\documentclass[oneside,12pt]{CUEDthesisPSnPDF}

\ifpdf

    \pdfcatalog { /PageMode (/UseOutlines)
                  /OpenAction (fitbh)  }
\fi

 \title{\textsf{\textbf{\normalsize{THE TWO-ENVELOPE PROBLEM: AN INFORMED CHOICE}}}}
 \collegeordept{\href{http://www.unisa.ac.za}{}}
 \university{\href{http://www.unisa.ac.za}{}}
 
\ifpdf

  \author{\href{mailto:jbtyler@nasor.co.za}{\textsf{\normalsize{by}}\\[8mm]\textsf{\normalsize{JEFFREY BRIAN TYLER}}}}

 \crest{\includegraphics[width=30mm]{unisa}}

\else

  \author{By\\JEFFREY BRIAN TYLER}

  \university{at the\\UNIVERSITY OF SOUTH AFRICA}

 \crest{\includegraphics[bb = 0 0 292 336, width=30mm]{unisa}}
 
\fi

\renewcommand{\submittedtext}{\textsf{\normalsize{submitted in accordance with the requirements for the degree of\\[10mm]}}}

\degree{\textup{\textsf{\normalsize{MASTER OF SCIENCE}}}}
\subject{\textsf{\normalsize{\textup{OPERATIONS RESEARCH}}}\\[8mm]\textsf{\normalsize{\textup{at the}}}\\[7mm]\textsf{\normalsize{\textup{UNIVERSITY OF SOUTH AFRICA}}}}

\degreedate{\textsf{\normalsize{SUPERVISOR: PROF P H POTGIETER}}\\\textsf{\normalsize{CO-SUPERVISOR: PROF G DAVIE}}\\[7mm]\textsf{\normalsize{MARCH 2014}}}

\usepackage{watermark}

\usepackage{booktabs}
\usepackage{ctable}
\usepackage{colortbl}
\usepackage{longtable}

\usepackage[centertags]{amsmath}
\usepackage{mathtools}

\usepackage{amsfonts}
\usepackage{amssymb,fullpage}

\usepackage{graphicx}
\usepackage{caption}
\usepackage{natbib}
\usepackage{pgfplots}

\usepackage{hyperref}
\hypersetup{colorlinks=false}

\usepackage{nomencl}
\makenomenclature
\makeindex

\usepackage{rotating}

\usepackage{calc}

\usepackage{tikz}
\usetikzlibrary{shapes, shapes.geometric, shapes.symbols, shapes.arrows, shapes.multipart, shapes.callouts, shapes.misc,positioning}

\usepackage{verbatim}

\usepackage{amsthm}
\usepackage{newlfont}

\usepackage[LGR,T1]{fontenc}

\usepackage{amsmath}
\usepackage{mathrsfs}
\usepackage[titletoc,toc,title,page]{appendix}

\usepackage{csquotes}

\newtheorem{proposition}{Proposition}
\newtheorem{experiment}{Experiment}
\newtheorem{possibility}{Possibility}
\newtheorem{rules}{Rule}
\newtheorem{case}{Case}
\newtheorem{interval}{Interval}

\begin{document}




\renewcommand{\appendix}{%
   \par
   \setcounter{section}{0}%
   \renewcommand{\thesection}{\thechapter\Alph{section}}%
}




\maketitle


\setcounter{secnumdepth}{3}
\setcounter{tocdepth}{3}

\setcitestyle{notesep={:},round,aysep={},yysep={;}}

\frontmatter 

\pagenumbering{roman}



\begin{dedication} 

Mathematics is the DNA of all that matters.

\end{dedication}




\begin{acknowledgements}      

\doublespacing

Professor Petrus Potgieter and Dr George Davie never gave up. In all my thinking and doing, Petrus and George believed in me and gave me every opportunity to realise my long held dream. Their support and understanding went beyond the academic realm, and for that, I am deeply grateful. Their unwavering attention, guidance, and enthusiasm sustained me through an arduous journey and I could not have asked for better companions. Petrus and George, I thank you.  

\end{acknowledgements}





\begin{abstracts}        

\doublespacing

The host of a game presents two indistinguishable envelopes to an agent. One of the envelopes is randomly selected and allocated to the agent. The agent is informed that the monetary content of one of the envelopes is twice that of the other. The dilemma is under which conditions it would be beneficial to switch the allocated envelope for the complementary one. The objective of his or her envelope-switching strategy is to determine the benefit of switching the allocated envelope and its content for the expected content of the complementary envelope. 

\bigskip

The agent, upon revealing the content of the allocated envelope, must consider the events that are likely to have taken place as a result of the host's activities. The preceding approach is in stark contrast to considering the agent's reasoning for a particular outcome that seeks to derive a strategy based on the relative contents of the presented envelopes. However, it is the former reasoning that seeks to identify what the initial amounts could have been, as a result of the observed amount, that facilitates the identification of an appropriate switching strategy. 

\bigskip

Knowledge of the content and allocation process is essential for the agent to derive a successful switching strategy, as is the distribution function from which the host sampled the initial amount that is assigned to the first envelope. 

\medskip

For every play of the game, once the agent is afforded the opportunity of sighting the content of the randomly allocated envelope, he or she can determine the expected benefit of switching.  

\end{abstracts}




\tableofcontents


\addcontentsline{toc}{chapter}{Nomenclature}


\def\baselinestretch{1}

\chapter*{Nomenclature}
\label{chap:nmncltr}

%
%

%
%

\begin{longtable}[l]{p{75pt} p{325pt}}
$X_1$ & The random variable that represents the initial amount of money assigned to the first envelope\\[2ex]
$X_1^\prime$ & The random variable that represents the amount of money assigned to the second envelope\\[2ex]
$X_2$ & The random variable that represents the toss of an unbiased coin\\[2ex]
$X_3$ & The random variable that represents the toss of an unbiased coin\\[2ex]
$Y$ & The random variable that represents the amount of money $y$ in the randomly selected and then allocated to the agent (player of the game)\\[2ex]
$Z$ & The random variable that represents the amount of money in the complementary envelope\\[2ex]
$\textbf{A}$ & The allocation matrix\\[2ex]
$\textbf{C}$ & The matrix representing the contents of the allocated and complementary envelopes\\[2ex]
$B$ & The random variable that represents the benefit (difference) between the content $z$ of the complementary envelope $Z$ and the content $y$ of the allocated envelope $Y$\\[2ex]
$E(B \big\vert Y = y)$ & The expected benefit when switching the amount $Y=y$ in the allocated envelope for the expected content of the complementary envelope\\[2ex]
$(\Omega,\mathcal{F},P)$ & The probability space\\[2ex]
$\Omega$ & The sample space\\[2ex]
$\mathcal{F}$ & The event space\\[2ex]
$\mathcal{C}_y$ & The set of events for consideration, by the agent, when observing the amount $y$ in the allocated envelope $Y$\\[2ex]
$C_y$ & A subset of $\mathcal{C}_y$\\[2ex]
$c$ & The event $c \in C_y$\\[2ex] 
$b(c)$ & The benefit of switching for the event $c$\\[2ex]
$e(y)$ & The exchange condition necessary for switching the allocated envelope for the complementary envelope\\[2ex]
$P(x)$ & Equivalently, $P(X=x)$\\[2ex]
$F_X(x)$ & The cumulative distribution function of the random variable $X$\\[2ex]
$f_X(x)$ & The probability density function associated with $F_X(x)$\\[2ex]
$X \sim f_X(x)$ & The random variable $X$ is distributed according to the probability density function $f_X(x)$\\[2ex]
$s(y)$ & The strategy associated with the observed amount $y$\\[2ex]
$\mathcal{G}$ & The {sub-$\sigma$-algebra} induced by $Y$\\[2ex]
$g(\omega)$ & A function that defines the factor applied to the initial amount to determine the amount assigned to the second envelope
\end{longtable}


\mainmatter 

\printnomenclature


\def\baselinestretch{1}

\chapter{Introduction}
\label{chptr:ntrdctn}

\begin{quote}\small

$\ldots$ statistical information on frequencies within a large, repetitive class of events is strictly irrelevant to a decision whose outcome depends on a single trial.

\begin{flushright}
--- Risk, Ambiguity, and the Savage Axioms \citep[2]{ED1961}
\end{flushright}

\end{quote}

\doublespacing
 
The two-envelope problem associated with the game of switching envelopes, which is responsible for the so-called \enquote{two-envelope paradox}, relies on the choices that confront an agent who must derive a strategy for retaining the randomly allocated envelope or switching envelopes and accepting the complementary envelope.

\medskip
 
The simplicity of the game belies the unprecedented amount of scholarly interest and attention it has garnered since first considered by \citet[p133]{KM1943}.  A number of papers cite both the origination of the problem and the related paradox in a variant of the \enquote{necktie} problem first posed by \citet[p133]{KM1943}. The game has since mutated into the current two-envelope problem, which has enjoyed notoriety because of its paradoxical interpretation in that it is alleged that it is always beneficial to switch envelopes irrespective of the content. 

\medskip

The \enquote{two-envelope paradox}, which has been the cause of much debate and consternation within the academic fraternity, centres on the strategy of an agent (player of the game) and rarely the related activities of the host (of the game).  

\medskip

While the two-envelope problem that has given rise to the paradox is reasonably simple in its presentation, its resolution and explanation have been somewhat elusive and, in some cases, unconvincing because most authors have failed to formulate the problem correctly. What is important, however, is that the agent can use the information he or she possesses in order to adopt a game strategy that allows him or her to determine the expected content of the complementary envelope upon sighting the content of the allocated envelope. 

\medskip
 
The objective of any strategy is to identify the benefit of switching the randomly allocated envelope and its observed contents for the expected content of the complementary envelope. It is very tempting to claim that the relative and therefore average content of the complementary envelope must be greater than the observed amount in the randomly allocated envelope for switching to take place. 

\medskip

Of the two primed envelopes, one is randomly selected and allocated to the agent. The agent, who has been informed that the content of one of the envelopes is twice that of the other, is then given the option of retaining the allocated envelope or switching envelopes in order to secure the envelope that contains the larger amount. 

\medskip

In addition to being informed that the content of one of the envelopes is twice the content of the other, the agent must also consider whether having sight of the content of the allocated envelope is relevant. If sighting the content is deemed to be significant, then it becomes important to determine what other information would assist the agent in identifying the game strategy that can realise the greatest expected benefit for any observed amount in the allocated envelope. Although \enquote{twice the other} seems to be a simple statement, this result can derive from halving, doubling, or a combination of halving and doubling, depending on the content and allocation process.

\medskip

In most previous articles, the authors assume that the agent will reason that if the randomly allocated envelope is observed to contain an amount of money, then the complementary envelope must contain either half or double that amount, each with an equal likelihood. However, it is not the content of the complementary envelope that is relevant. It will be realised that it is the content of the first envelope and the consequent content of the second envelope that are directly related to the observed amount in the allocated envelope. And, by inference, indirectly the content of the complementary envelope is considered relevant, when in reality it is not. 

\section{Information}

The host of the game is responsible for assigning amounts of money, according to a certain content and allocation process, to each of the two envelopes. How these amounts are identified and assigned to the envelopes is explained in the content and allocation process of the game. The money assigned to the first envelope is a randomly determined amount, whereas the second amount is dependent upon the amount assigned to the first envelope. 
\medskip

In the random allocation of these indistinguishable envelopes, the agent receives one of the two envelopes.
  
\medskip

Newcomers to the two-envelope problem may be forgiven for an initial response that has been exhaustively researched and, depending on the objective of the study in question, justified. In most instances, such examinations have ignored the activities of the host --- the identification of the initial amount that is to be assigned to the first envelope and the subsequent amount assigned to the second envelope --- for a specific play of the game, with the presented envelopes and their contents treated as the only relevant data for the agent. Indeed, it has been assumed that the outcomes of the host's activities for each play of the game are the only factors deemed to be relevant for solving the problem. 

\medskip

If the agent considers the problem from a probabilistic perspective, it soon becomes apparent that the activities of the host --- for all related outcomes --- directly influences the agent's potential strategy. It also becomes clear that the observed amount may be related to the activities of the host on an expected basis and that there is an indirect relationship between the amounts outside of the immediately obvious linkage. Specifically, what the agent reveals in the allocated envelope does define the possible activities of the host (who was responsible for the initial amount assigned to the first envelope and consequent amount assigned to the second envelope).

\medskip

This problem can be considered to be a two-stage experiment. The first stage is associated with the activities of the host,  while the second stage is associated with the activities of the agent.  Because these two stages are subtly independent, the agent must consider the following five content and allocation possibilities (one of which will have been adopted by the host) in order to derive his or her switching strategy. 

\subsection{Content and allocation}

The initial amount is determined by the host who randomly samples from an appropriate cumulative distribution function and assigns the amount to the first envelope. Thereafter, the second amount is derived according to one of the following five possibilities:

\begin{possibility}\label{pssblty:one}
The host may prime the second envelope, by halving or doubling the initial amount of money, and assigning the resultant amount to the second envelope. He or she then randomly selects one of the envelopes and allocates it to the agent for consideration;
\end{possibility}

\begin{possibility}\label{pssblty:two}
The host may prime the second envelope, by doubling the content of the first. He or she then randomly allocates one of the envelopes to the agent for consideration;
\end{possibility}

\begin{possibility}\label{pssblty:five}
The host may prime the second envelope, by halving the content of the first. He or she then randomly allocates one of the envelopes to the agent for consideration;
\end{possibility}

\medskip

While it is true that the content of the complementary envelope is either half or double the content of the randomly allocated envelope, it is true that the expected benefit of switching envelopes is the difference between the expectation of the content of the complementary envelope and the observed content of the allocated envelope. 

\medskip

In order to identify an appropriate switching strategy, it is thus relevant for the agent to consider what the host may have assigned initially to the first envelope and subsequently to the second envelope. The expected benefit of switching, which is a function of the amount observed by the agent, is associated with the host having assigned either half the amount, the equivalent amount, or twice the observed amount to the first envelope. And, the content of the second envelope is now considered as a function of the content of the first envelope. This, allocation of the content to the first and envelopes is not the end of the story. The envelopes are indistinguishable and subject to a random allocation that will see different combinations that ultimately generate the allocated and complementary envelopes. Since the host is randomly selecting an initial amount from an arbitrary distribution function it is distinctly possible that each of the amounts -- half, equivalent, or double -- will each occur with a different probability. Moreover, it is accepted that one of the indistinguishable envelopes is randomly allocated to the agent and that the complementary envelope will be considered to be preferred or not upon sighting the content of the allocated envelope, when permitted. 

\medskip 

It is the random allocation of the envelopes that ensures that the allocated envelope contains a fixed amount and that the complementary envelope will contain either half or double the content of the allocated envelope. That the content of the complementary envelope contains half or double that of the allocated envelope agrees with the relative explanation that is so often quoted. 

\medskip

\begin{possibility}\label{pssblty:three}
The host may allocate the first envelope to the agent and then prime the second envelope by either halving or doubling the content of the first; or
\end{possibility}

\begin{possibility}\label{pssblty:four}
The host may retain the first envelope. The second envelope will  then be primed by either halving or doubling the content of the first envelope. This second envelope is then allocated to the agent.
\end{possibility}

\medskip

The traditional approach to solving the two-envelope problem has only considered the content $Y$ of the allocated and the content $Z$ of the complementary envelope. For any play of the game it will be taken that the contents of the envelopes remain unchanged. With this condition it must be accepted that, irrespective of the content and allocation process, that the relative content of the complementary envelope is either twice or half that of allocated envelope. For any specific play of the game (single trial) the allocated envelope may contain the larger amount $Y = 2x$ and the complementary envelope $Z=x$, and as a result $Z = Y/2$. Alternatively, if the allocated envelope contains the smaller amount, then $Y=x$ and $Z=2x$, and as a result $Z=2Y$. Which transformation should be applied is clearly an issue that cannot be ignored. In the absence of information the agent can conclude that the host primed the envelopes with the amounts $X_1 = y$ and $X_1^\prime =2y$, if the \enquote{doubling-only} process was adopted. Therefore, as a result of the random allocation, the agent may observe $Y=y$ and consequently conclude that the complementary envelope contains $Z=2y$ or the agent may observe $Y= 2y$ and consequently conclude that the complementary envelope contains $Z=y$. Since these outcomes are equally likely -- due to symmetry -- the agent should be indifferent to switching envelopes or not.

\medskip
It would be more appropriate to take the content of the allocated envelope $Y=y$ as fixed and to then consider the possible values that the complementary envelope $Z$ may contain as a result of the content and allocation process. While it is acknowledged that the relative content of the complementary envelope $Z$ can be either $y/2$ or $2y$, these possible values should only be considered as a result of the content and allocation process. 

\medskip

The amount in the complementary envelope is assumed to be associated with the host generating an amount $X_1=y$ and does not recognise that the amount $Y=y$ can only be observed as a result of the host's activities associated with $x_1 \in \left\lbrace y/2;y;2y\right\rbrace$. Further, the assertion that $Y=y$ is only possible for $X_1=2y$ or $X_1=y/2$ does not consider the relationship between the observed $Y=y$ and $X_1$ and states that this relationship is distribution free by having originated from the host generating $X_1=2x_1$ or $X_1={x_1}/2$. The realisation that $X_1=2y$ and $X_1=y/2$ are just as relevant as $X_1 = y$ for the derivation of the expected benefit $B$ is discussed in Chapter \ref{chptr:strategies}, where the alternative strategies associated with the content and allocation process are derived. The evolution of the two-envelope problem and the consequent paradox is then discussed in Chapter \ref{chptr:paradox}.

\medskip

The way in which the problem is posed lends itself to assumptions about the reasoning of the agent. That the agent is assumed to reason in  a particular way is not necessarily correct. However, it does seem to be the approach followed by previous authors \citep{SJ1994}.

%
%
%
%
%

\subsection{The rules of the game}

While little or no attention has been paid to how the activities of the host influences the process of assigning the amounts to the envelopes and the subsequent allocation of the envelopes, it is important for the agent to know about these. This information must include the distribution function from which the host sampled the initial amount, while the agent must also have sight of the content of the allocated envelope.

\medskip

The rules of the game for the informed agent are thus as follows:

\begin{rules}\label{rule:one}
The agent will have sight of the contents in the allocated envelope;
\end{rules}

\begin{rules}\label{rule:two}
The agent will be informed about the content and allocation strategy;
\end{rules}

\begin{rules}\label{rule:three}
The agent will be informed about the process of amount identification;
\end{rules}

\begin{rules}\label{rule:four}
The agent will be informed about the distribution function from which the host sampled the initial amount. If the agent is not informed about the distribution function from which the initial amount is generated, the agent may assume a prior distribution. In addition, subsequent to each play of the game, the agent may refine the prior distribution to enhance the predictive ability in the successive play of the game. Irrespective of whether the agent has assumed or derived such a prior distribution or not, the density function under consideration for each play of the game will be taken as given;
\end{rules} 

\begin{rules}\label{rule:five}
The agent will be informed of the bounds (if they exist) for the amounts that can be assigned to the envelopes. While it is acknowledged that the paradox does not exist in the presence of a defined upper bound this still does not address a playing strategy.
\end{rules}

\medskip

The relevance of each of these rules is fundamental for solving the two-envelope problem.

\medskip

%
%
%

\section{Expectation}

The scholarly interest that the two-envelope paradox has generated is based on its simple presentation. But, it is the formulation of the problem (where the absence of information and as a result assumptions) that has perpetuated the assertion of the paradox.

\medskip

The traditional explanation of the paradox is founded on the premise that the outcome of any game is the difference between what is expected and what is observed. In particular, what is expected is taken as the average of two amounts, namely (i) double the observed amount and (ii) half the observed amount. 

\medskip

As described earlier, the host of the two-envelope game prepares two indistinguishable envelopes by assigning amounts of money to each. The agent is randomly allocated one of the envelopes. Before any interaction with the envelopes, the agent is informed that the amount of money in one envelope is twice the amount in the other. The agent is then asked whether in the absence of sighting the content, the complementary envelope would be preferred.

\medskip

However, nothing has been said about sighting the content. Herein, it is shown that in the absence of the distribution function from which the host identified the initial amount, there is no benefit for the agent to switching envelopes. It is only once the density function is considered that the agent can derive a switching strategy. 

\medskip

For many the approach is associated with the host will double the initial amount and assign this to the second envelope. After randomly allocating one of the indistinguishable envelopes to the agent, the agent can conclude that the envelope will contain an amount that could be the larger or the smaller. If the agent is informed that the host has adopted a doubling only process, then the agent can safely assume that if the envelope contains the larger amount then the initial amount assigned to the first envelope would have been the smaller amount --- a relative amount that is also stated as half the amount in the allocated envelope. The probability of the host generating this smaller amount would have been identified from the distribution function chosen by the host, and this smaller amount would have been doubled, hence the larger amount. Alternatively, if the envelope contains the smaller amount then the agent can conclude that the initial amount was assigned to the first envelope and that the larger (double the smaller) was assigned to the second envelope --- a relative amount that is also stated as twice the amount in the allocated envelope. The complementary envelope does, as noted by many, contain either half or double the amount in the allocated envelope. The content of the allocated envelope is fixed. Applying a probability to the respective initial amounts does yield the expected outcome and by subtracting the amount contained in the allocated envelope the expected benefit of switching can be derived.

%
%
%
%
%

\section{A strategy or a paradox}

Although two envelopes are under consideration in every play of the game, only the content of the randomly allocated envelope is relevant once its content has been revealed for any play. The content of the complementary envelope is irrelevant for the purposes of deriving a strategy unless the agent has not been given information of the prior distribution function from which the host is sampling (in which case the content of the second envelope can be utilised for updating the prior distribution). Hence, the agent can collate the outcomes for each play of the game in order to monitor the overall performance of the adopted strategy.

\medskip

The traditional posing of the problem has led to it being improperly formulated and consequently incorrectly solve \citep{SE1994}. This ill-stated problem, in the absence of any assumption and information, together with the failure to re-formulate the problem correctly is the reason for the plethora of papers that continue to sustain the myth of the paradox \citep{LT2004}.

\medskip

The argument, for the envelopes containing $X_1$ and $X_1^\prime$, in the absence of sighting the content, is that if the agent had been randomly allocated the envelope that contained the amount $Y = X_1$ then the benefit $B$ of switching would be the difference between $X_1^\prime - X_1$. Alternatively, if the allocated envelope contained the amount $Y = X_1^\prime$, then the benefit $B$ of switching would be the difference between $X_1 - X_1^\prime$. Each of the envelopes is equally likely to be allocated to the agent. The agent may assume that since the contents are fixed at the time of selection, they remain unchanged during the process; therefore, it is irrelevant whether he or she chooses to switch or not.

\medskip  

The alternative argument, in the absence of sighting the content, runs as follows. The complementary envelope contains either half or double the amount in the randomly allocated envelope for the following reason. If the content of the allocated envelope were observed to be $Y$, then the complementary envelope would be either $Z=2Y$ and the benefit of switching would be $B = Z - Y = Y$. Alternatively, if the complementary envelope were $Z=Y/2$, then the benefit of switching would be $B = Z - Y = -Y/2$. Since each of these outcomes is equally likely, the amount you stand to gain is greater then the amount that you will lose, and the agent should switch. But, herein lies the paradox. For if the complementary envelope is treated as the randomly allocated envelope then the same reasoning applies.  

\medskip

It is merely coincidental that without considering the activities of the host and the consequent statistical events, the conclusions of many papers concur fortuitously with the findings presented in the subsequent chapters of the present paper.

\medskip

In order for the agent of the two-envelope game to identify an envelope switching strategy, he or she must be informed of the host's activities, which are associated with the identification of the initial amount assigned to the first envelope.

\medskip

It is demonstrated herein that once the agent has been informed about the distribution function, and upon sighting the content of the allocated envelope, he or she can derive and apply a suitable strategy to identify the expected benefit of switching. Consequently, the agent can decide whether to switch given the available information. Examples of different density functions are presented in Chapter \ref{chptr:examples}.

%
%

\section{Summary}

The agent has to identify a game strategy that will realise the greatest expected benefit for each play of the game, either by retaining the randomly allocated envelope and its contents or by switching envelopes and accepting the content of the complementary envelope. The difference between the expected content of the complementary envelope and the observed content of the randomly allocated envelope is then the expected benefit of switching. 

\medskip

In the presence of the density function of the associated events for the observed amount, a benefit can be identified from switching the allocated envelope for the expected content of the complementary envelope. It is not the content of the complementary envelope, for any single trial, that dictates the strategy. It is the expected content of the complementary envelope, for which the content of the allocated envelope is always noted to be $Y=y$, that generates the necessary switching strategy. The expected content of the complementary envelope is directly related to the initial amount assigned to the first envelope.

\medskip

The agent, however, in revealing the content of the randomly assigned envelope, no longer needs to be concerned with the content of the complementary envelope. The sighted content is the means whereby the expected content of the complementary envelope is derived.

\medskip

When the agent is not informed about the density function, he or she can use the information that each play of the game presents in order to refine an assumed prior distribution, which would be useful in subsequent plays of the game. Therefore, the approach adopted throughout this paper relies on the fact that a density function, whether derived or provided, is given.

\medskip

In deriving his or her strategy, the agent considers the content of the randomly allocated envelope and the expected content of its complement. The expected difference between the contents of the former and the latter expresses the benefit to be realised should the agent decide to make this switch. 

\medskip 

Thus, the agent assesses the available information associated with the content and allocation processes before adopting his or her chosen strategy.  

\medskip

Information on the activities associated with the host's generating of the game is essential for the agent to identify an appropriate and optimal playing strategy. Further, the agent is assumed to be risk-neutral and rational.

\medskip

In summary, this research shows that, in the presence of information, it is possible for the agent to determine a switching strategy that can identify the expected benefit of switching envelopes for any observed amount in the randomly allocated envelope. Even after the play, the gained knowledge of the content of the envelopes has no bearing on successive plays of the game if the agent is informed about the distribution function from which the host sampled the initial amount that was assigned to the first envelope.




\chapter{Literature Review}

\begin{quote}\small

You know that envelope A contains a randomly determined sum of money, and envelope B contains, as a consequence of the flip of a fair coin, either half or twice that sum. And you know that one of these envelopes, randomly selected, has been given to you.

\medskip

Suppose you are a risk-neutral money seeker who can choose between the money contained in the envelope that has been given to you and the money in the other envelope. Which do you prefer, or are you indifferent?

\medskip

Suppose that you open your envelope. Now that you know exactly how much money is in it, which do you prefer, this money, or the money in the other envelope?

\begin{flushright}
--- Two Envelopes \citep[p1]{SJ1994}
\end{flushright}

\end{quote}

The above-described dilemma posed by \citet[p1]{SJ1994} is typical of the majority of papers that discuss the two-envelope problem, which as noted in Chapter~\ref{chptr:ntrdctn}, has engendered a great deal of research interest and debate. The challenge it has presented to many and the consequent explanations by authors have not necessarily produced a convincing argument to substantiate or refute the two-envelope paradox, however. 

\medskip

A story be told about an interesting paradox? But, perhaps even more interesting is the story that can be told of the many attempts to resolve an otherwise simple problem and thereby the paradox. That the paradox has attracted so much interest from such a diverse array of authors is a testament to the intrigue surrounding the problem and possibly academic prowess. 

\medskip

It is true that the amount of money in one envelope is twice that in the other envelope. But what is the implication of this statement that would seem to be quoted in nearly every publication. What is the true meaning of this statement and how can the other envelope always contain a greater expected amount of money regardless of which envelope you choose?

\medskip

And it can be said that the observed amount is possibly the smaller or the larger of the content of the two envelopes. It can also be said that the content of the one envelope is twice the other, or half the other. These innocent statements only obfuscate what is the essential for the resolution of this problem.

\medskip

What is necessary is that the reader recognises the provided information and then proceeds to understand the activities of the host of the game. In so doing, the agent will come to realise what additional information would make it essential for a winning strategy. 

\medskip

That the problem is ill-posed has been recognised by \citet[p2]{LT2004} and a re-formulation is necessary to facilitate the derivation of an appropriate switching strategy.

\medskip

The alert reader will note the lack of consistency and the many attempts from diverse disciplines of those that have sought to settle the issue. We should heed the warning by \citet[p5]{CC1995} of the the subtle nature of the problem and exercise caution. Many are tempted to resolve the two-envelope problem through all manner of means when relying on increasingly complex arguments and may miss the point \citep[p97]{RP1994}. 

%
%
%
%

\section{Posing the problem}

The variants, of which this version by \citet[p1]{SJ1994} is but one, are essentially all the same. This is simply a mathematical problem that is not well-posed and over time many authors have recognised this and consequently re-formulated the problem and made it more tractable \citep[p19]{SE1994}.  

\medskip

The presentation of the two-envelope problem rarely informs the agent about the content and allocation process that is adopted by the host. There is also some degree of confusion about the distinct difference between the amounts assigned to the first and second envelopes and what is contained in the allocated and complementary envelopes \citep[p98]{RP1994}.  

\medskip

Recent papers have referred to the density function from which the host selects the initial amount of money. If the agent is informed about this density function and understands the content and allocation process adopted by the host, then he or she can identify a strategy, provided that the content of the allocated envelope is sighted before deciding to switch envelopes or not.  

\medskip

Most authors in their exploration of the problem consider the amount of money in the allocated envelope and conclude that the amount in the complementary envelope is directly derivable from this sighted amount.

\medskip

The real issue is being able to identify the amounts in the first and second envelopes that would have yielded the observed amount in the allocated envelope. It is the initial amount assigned to the first envelope and the consequent assigning of another amount to the second envelope followed by the random allocation of the indistinguishable envelopes that determines what the agent reveals in the allocated envelope.

\medskip

%
%
%

\subsection{The formulation}
 
While, not being permitted to sight the content may be taken as information, it does render the decision to switch or not irrelevant and consequently the agent should be indifferent. 

\medskip

The realisation that in the absence of sighting the content of the allocated envelope there can be no benefit to switching the envelopes has been thoroughly reported by \citet[p2]{LT2004}. But, it is the alternative situation, when the agent is afforded the opportunity of sighting the content of the allocated envelope that continues to frustrate and entertain \citep{BJ1995, CS2000}. 

\medskip

Once the host informs the agent of the content and allocation process, then the agent can derive a strategy. In the instance where the host withholds the information of the distribution function from which the initial amount would have been identified, then the agent can in the long run estimate and refine the prior distribution subsequent to each play of the game. As a result of this, it will be taken that the agent is informed of the distribution function. Strategies, in instances of the agent being informed, are readily identified by \citep[p275]{CR1992}, and \citep{BK1995, BR2000, SS1997, BJ1995, CB1994}. 

\medskip

It is not the expectation of the density function that is in question but the existence of the {sub-$\sigma$-algebra} induced by the observed amount \citep[p254]{AR1972}.  Once this is accepted, then the expected benefit of switching is the conditional expectation of the {sub-$\sigma$-algebra} induced by the observed amount. Any question of infinities can immediately be dismissed, and any concerns about the measures associated with contentious distribution functions are no longer valid \citep[p176]{NB1989}, \citep[p98]{RP1994}.

\medskip

The concept of a {sub-$\sigma$-algebra} ensures that the observed amount in the allocated envelope, also referred by \citet[p100]{RP1994}, \citet[p8]{JR1986}, and \citet{MC1997} as a rigid designator, will allow the narrowing of the distance between opinions. Finally the consensus -- albeit from different disciplines -- will add credence to the overwhelming evidence refuting the existence of the paradox \citep[p101]{RP1994}.

%
%

\subsection{The content}

What is assigned to the first and second envelope must be distinguished from what is in the allocated and complementary envelopes. 

\medskip

Most resort to describing the problem as consisting of two envelopes to which have been assigned sums of money. In the problem posed by \citet[p1]{SJ1994} above, the amount assigned to the second envelope is either half or double the amount assigned to the first. This content and allocation process is also considered by \citet[p212]{CJ1992}, \citet{NB1989}, \citet{BF1996}, \citet[p97]{RP1994}. By contrast \citet{BJ1995} considers only that the amount assigned to the second envelope is double the amount assigned to the first envelope. While \citet{BK1995} assign an amount to the second envelope that is half the amount assigned to the first envelope. These different processes, of content and allocation, for the two-envelope problem yield distinctly different strategies, as will be demonstrated in Chapter~\ref{chptr:strategies}. 

\medskip

The allocation strategy will be discussed in Chapter~\ref{chptr:allocation}.

\medskip

What process is responsible for observing an amount and then concluding that the other is half or double, is rarely considered. It is routinely taken that the  relevant information is immediately before the agent and is all that is necessary to justify an appropriate strategy. Although the agent observes an amount of money, the amount observed is not what the host may have assigned to the first envelope. And this realisation makes it necessary for the agent to know the distribution function from which the initial amount was sampled \citep[p5]{LT2004} . For example, in the instance of the \enquote{doubling-only} process, for the agent to observe a particular amount of money, it is possible that the host may have sampled an initial amount that is exactly what the agent observes and the complementary envelope then contains double the observed amount. Or, the initial amount could have been half the amount observed and as a result of the \enquote{doubling-only} process the amount assigned to the second envelope is revealed as the content of the allocated envelope. This later alternative ensures that the complementary envelope contains half the the amount in the allocated envelope. 

\bigskip

What matters is the content and allocation process and what the agent observes in the allocated envelope. With this in mind, the content of the complementary envelope at the time of play, is irrelevant to the agent. 
\medskip 

The approach in this paper follows the discussion by \citet[p914]{KO2007} in which they recognise that the amounts of money assigned to the envelopes, according to the content and allocation process, are relevant for the agent to derive an appropriate switching strategy.

\begin{quote}\small

We turn, now, to the fixed-sum formulation of the two envelope paradox, where the two mutually exclusive and jointly exhaustive possible states of the world that determine the consequences of your actions are characterised in terms of the actual contents of one of the envelopes, say A. \citep[p919]{KO2007}

\end{quote}

\bigskip

Traditionally, in the version of the two-envelope problem where the agent is afforded the opportunity of sighting the content of the allocated envelope, the probability associated with the content being either the larger or the smaller is taken as the same. However, the probability of the allocated envelope containing the larger amount may not be the same as the probability of the envelope containing the smaller amount. Even considering this possibility detracts from the problem and in this paper it will be demonstrated that what one observes in the allocated envelope is a function of what was assigned to the first envelope and the issue is not whether the allocated envelope contains the smaller or the larger amount. It is the process adopted by the host that dictates what the agent will observe, in the allocated envelope, that is now relevant. Is the envelope into which the initial amount assigned allocated to the agent before deciding what is to be assigned to the second envelope? In this instance, the \enquote{halving or doubling} method must be used, since if the \enquote{doubling-only} method were adopted, then the agent would know that the complementary envelope contained twice the content of the randomly allocated envelope, and it would be certain in all plays of the game that a switch would result in a guaranteed benefit.

\medskip

It will be demonstrated in Chapter ~\ref{chptr:allocation} and Chapter ~\ref{chptr:strategies} that while it is true that the content of the complementary envelope is twice or half of the content of the allocated envelope, it is nonetheless meaningless and of no relevance to the resolution of the problem. 

\medskip

%
%
%

\section{The method}

It is also true that once the first and second envelopes are primed that each have equal likelihood of been allocated to the agent. But, as will be noted in the strategy chapter such probabilities are factored out in the expected benefit functions. The relevance of equal likelihood for the allocation of the two envelopes is questioned by \citep[p92]{AK2005}

\medskip

The contextualising of the dilemma in the form of a story necessitates that the reader must distill from the verbosity the essentials in order to formulate the problem. \citet{SR1992} tackles the problem by not embracing probability. A subtle wording of what can be gained or lost by switching would appear to be quite convincing in his particular argument. Because the alternatives are equally likely, probability should be considered \citep{CJ2002, CC1995}.

\subsection{Reason and logic}

 Some authors such as \citet{SR1992} and \citet{CC1995}, pose the problem as a story and then proceed logically to solve it. 

\medskip

\citet{CJ2007} and \citet[p102]{RP1994} argue that neither Proposition \ref{prpstn:one} nor Proposition \ref{prpstn:two}, as proposed by \citet[p190]{SR1992}, are correct, and the very wording of the propositions is questioned by \citep[p92]{AK2005}. 

\medskip

It is acknowledged by many authors that no resolution to the paradox is possible with the available information. What is necessary is additional information, or the agent will need to make assumptions before attempting to resolve the problem. As such the problem is open to interpretation and the vague statement of the problem is probably the very reason that the literature is peppered with such diverse responses.
 
\medskip
 
Some authors \citet{CJ1992}, \citet{SR1992} and \citet{CJ2007} tackle the problem using disciplines outside of the field of probability. Some papers \citep[p10]{SP2010} of a philosophical nature acknowledge the need to resolve the problem by embracing a probabilistic argument. It will be noted that there is a convergence of argument and in many instances there is an agreement albeit from different disciplines. This confluence of ideas and argument are all lending credence to the demise of the apparent paradox. 

%
%
%
%

\subsection{Probability}

The many different attempts to explain the legitimacy of the paradox stems from the fact that many authors are not well versed in the discipline of probability. The fundamentals essential to the probabilistic approach would seem to be taken for granted and in many instances the mere referencing to Bayesian or conditional probabilities would seem to add an air of legitimacy to the authors intent. The narrative that is used to deliver the problem is not very clear and care, as advocated by \citet[p2]{VR2012}, needs to be exercised by the reader who is tempted to claim that he or she can resolve the two-envelope problem and thereby explain the paradox. In many instances attempts to resolve the paradox only reinforces and perpetuates the myth.

\medskip

According to \citet{BJ1995, JF1994}, the distribution from which the host samples the initial amount is essential for the agent to derive an appropriate switching strategy. The distribution function that represents the initial monetary amount from which the host will sample, can be either discrete or continuous; this however is never mentioned in the usual statement of the problem. As shown in Chapter~\ref{chptr:strategies}, this information is essential for the agent to derive an appropriate density function. For any successful strategy, the density function must be known by the agent to support the identification of an appropriate switching strategy \citep{JF1994}. 

\medskip

While many authors recognise the conditional probability (Bayesian) is necessary to derive an appropriate switching strategy, they unfortunately fail to recognise that the expected benefit is conditional on the {sub-$\sigma$-algebra} induced by the amount observed in the allocated envelope \citep[p249]{AR1972}. This necessitates that the agent be informed of the distribution function from which the host sampled the initial amount that was assigned to the first envelope. The resultant event space immediately lends itself readily to a solution that overcomes many of the concerns that have been expressed, as a means to justify or dismiss the paradox \citep[p12]{NY1234}. Concerns, about infinite amounts of money that are labelled as absurd and devoid of all reality and therefore dismissed are no longer an issue when the event space is recognised as a {sub-$\sigma$-algebra}. Whether the amounts of money in the financial system are infinite - the FED can always print more - or not, is besides the point \citep{RP1994}. 
 
\medskip

The expected benefit is succinctly expressed as the conditional expectation for the {sub-$\sigma$-algebra} induced by the observed amount, as discussed by \citet[p3]{SA2013} and \citet[p249]{AR1972}.

%
%
%

\section{Switching strategies}

In instances where the amounts are bounded, strategies can be derived without much effort, as will be detailed in Chapter~\ref{chptr:strategies}.

\medskip

It can be said that whether a distribution function has a finite mean or not is irrelevant for the identification of an appropriate switching strategy. \citet{BJ1995} correctly states the problem that the agent is confronted with as one where he or she must consider the expected content of the complementary envelope once the content of the allocated envelope is revealed.
 
\medskip

In some instances, the density function may support the notion that switching is always beneficial. In no way is this paradoxical; it is rather a consequence of the density function. Any density functions with a finite mean will not support the paradox. \citet[p8]{BJ1995} states that a switching strategy can be identified that will yield the greatest expected benefit for any observed amount from the allocated envelope. While distribution functions possessing no finite mean will always support the paradox, they render sighting the content of the allocated envelope redundant and support the notion that switching is always beneficial.

\medskip

If the amounts assigned to the envelopes by the host are bounded, then knowledge of this by the agent is necessary. In such an instance, the paradox would not exist since it would not be beneficial for the agent to switch if the content of the allocated envelope were found to be greater than half the upper bound. Should the observed amount be less than twice the lower bound, however, then it would be advantageous to switch.

\medskip

It is possible for the agent, upon sighting the content and observing the amount $Y=y$, to determine the conditional expectation of the benefit when switching envelopes. It will be realised that the conditional expectation is a function of the {sub-$\sigma$-algebra} induced by $Y$. This realisation immediately eliminates some of the absurdities that have been espoused as the cause of the paradox. Absurdities such as the infinite sets, are immediately reduced to finite sets, as a result of the induced {sub-$\sigma$-algebra} \citep{SA2013}. 

 \medskip
 
It has been proposed by \citet[p3,p12]{MA2009}, \citet[p7]{BR2000}, and \citet[p14]{GA2011} that a strategy could be based on \enquote{Cover's switching function}. This will be less than ideal in some instances and an exact switching strategy can be derived from the provided distribution function. In the absence of a provided distribution function, through an adaptive strategy, the agent can assume a distribution whose parameters are refined upon the completion of each game. 
 
 \medskip

It is also worth mentioning, as proved by \citet[p10]{BJ1995}, that no distribution function possessing a finite mean can be \enquote{paradoxical}. In such instances there will be an interval over which it would be beneficial to switch envelopes and alternative intervals where switching would not be beneficial. Similarly, in the presence of bounds, it will not always be beneficial to switch.

\subsection{The paradox}

Most authors attempt to resolve the paradox by comparing the content of the two envelopes. For whatever the content of the envelopes, the complementary envelope will always be half or twice the content of the allocated envelope. This justifies the switching. In this paper the content of the allocated envelope is treated as a constant (rigid designator) and the possibility of the content of the allocated envelope taking on the larger or the smaller amount is never in doubt. As such this aligns with the philosophical approach proposed by authors \citet{MC1997}, \citet{RP1994}, and \citet{PR2007}, who claim to have settled the issue of the paradox. In most of these instances they are dismiss the paradox on the grounds that the content of the allocated envelope must treated as a rigid designator \citep{CJ2007,SP2010,CJ2002}. These arguments call on the disciplines of logic and are in this paper substantiated by the appropriate application of probabilistic argument.

\medskip

\citet{SP2010} recognises the need for a probabilistic approach and that the inevitable resolution of the two-envelope problem will be understood by philosophers.

\medskip

\citet{BJ1995} approaches the paradox by considering that the smaller monetary amount is essential and then proceeds to derive a function that expresses the expectation as a result of switching. There are two distributions presented by \citet{BJ1995} to support the strategy of always switching. \citet{BJ1995} acknowledges that these distributions, although valid, possess no finite mean, and demonstrates that no distribution with a finite mean (even an infinite distribution) can be paradoxical.
 
%
%
%

\section{Summary}
 
As will be noted from the aforementioned there are any number of attempts to resolve the two-envelope problem from a diverse array of disciplines. Many practitioners have relied on an extensive array of simple and complex arguments to justify their conclusions. And in most instances there is a reliance on a mathematical formulation of the problem. It is this formulation that I believe is the cause of the paradox. As will be seen, provided that there is information, then a strategy can be identified that will yield a greatest expected benefit for  any play of the game.

%
%
%
%



\def\baselinestretch{1}

\chapter{Posing the {two-envelope} problem}
\label{chptr:allocation}

\doublespacing

\begin{quote}\small

What is essential is invisible to the eye.

\begin{flushright}
--- The Little Prince \citep[p70]{AS1943}
\end{flushright}

\end{quote}

The game of switching envelopes, which lends its name to the two-envelope paradox, is based on the problem that confronts an agent who must identify a strategy to maximise the expected benefit by either retaining the allocated envelope and its contents or choosing the complementary envelope and its contents. The host of the game is responsible for assigning the amounts of money, according to a specified content and allocation process, to each of the two envelopes. The money assigned to the first envelope is a randomly identified amount that is distributed according to some probability density function. The second amount is dependent upon the amount assigned to the first envelope; it could be either half or double that amount depending on the \enquote{content} component of the content and allocation process.

\medskip

The random allocation of the two indistinguishable envelopes will see the agent receiving one of them. Neither the agent nor the host at the time of allocation can discern the contents and neither would they, until the contents are revealed, be in a position to comment on the appropriate switching strategy.  

\medskip

We divide the  two-envelope game into two stages: the activities of the host (stage 1) and the strategy of the agent (stage 2). The host of the game conducts three experiments in stage 1. The first experiment is the random sampling from an arbitrary distribution function, the second experiment is the tossing of an unbiased coin, and the third and final experiment is the tossing of the same coin again. In stage 2, the agent identifies the envelope-switching strategy that yields the greatest expected benefit, which is determined by considering the expected difference between the content of the complementary envelope and the content of the allocated envelope.

\medskip

\citet{BR2000} and \citet{BJ1995} are among the authors to state that the content (the smaller amount) of the first envelope is identified by the host  and that a subsequent amount that is twice the initial amount is assigned to a second envelope. On the other hand, \citet[p69]{SJ1994} refers specifically to halving or doubling the initial amount depending on the outcome of a random event. In yet other instances, no consideration is paid to the content and allocation process. However, this paper demonstrates that distinguishing between the two processes is important as they may yield distinctly different strategies. Where sighting the content of the randomly allocated envelope is prohibited, the content allocation process is irrelevant \citep[p117]{BF1996}.

%
%
%

\section{The host's activities}

The concept of a probability space $(\Omega,\mathcal{F},P)$ is introduced, where $\Omega$ is the sample space, $\mathcal{F}$ the event space, and $P$ the probability measure on $\mathcal{F}$. All subsets of $\Omega$ are events, and $\mathcal{F}$ consists of all subsets of $\Omega$. 

\medskip

Experiment~\ref{xprmnt:one}, performed by the host, involves identifying the initial amount of money that is assigned to the first envelope. The initial amount is drawn from  a cumulative distribution function, which is either given or derived from an arbitrary probability density function. 

\medskip

The outcome of Experiment~\ref{xprmnt:two}, performed by the host, is used to decide whether to halve or double the initial amount and assign the resultant amount to the second envelope.

\medskip

The outcome of Experiment~\ref{xprmnt:three}, performed by the host, is used to decide whether to allocate the first or the second envelope to the agent. The remaining envelope will be referred to as the complementary envelope.

\medskip

The grand experiment, for consideration by the agent, and the associated product probability space is a function of these three independent host-conducted experiments. Associated with each of them is an appropriate probability space that would need careful consideration by the agent in order for him or her to identify an appropriate strategy.  

%
%
%

\begin{experiment}\label{xprmnt:one}

$X_1$ is a random variable on $(\Omega_1; \mathcal{F}_1;P_1)$. $X_1$ represents the initial amount of money identified by the host and assigned to the first of the two envelopes. 

\medskip

The sample space associated with the drawing of an initial random amount $x_1$ represented by the random variable $X_1$, where $X_1 \sim f_{X_1}(x_1)$, is defined as:

$$\Omega_1 = \left\lbrace \omega_1\colon x_l < X_1(\omega_1) \le x_u \right\rbrace.$$

This initial amount $X_1(\omega_1) =  x_1$ is assigned to the first of the two envelopes. Unless otherwise specified, the notation $X_1(\omega) = x_1$ is hereafter represented by $X_1 = x_1$. The sample space for drawing the initial amount is determined by the host. We can see it as the values between $x_l$ and $x_u$.

\end{experiment}

\begin{experiment}\label{xprmnt:two}

$X_2$ is a random variable on $(\Omega_2; \mathcal{F}_2;P_2)$. $X_2$ represents the outcome of the toss of an unbiased coin. The sample space is

$$\Omega_2 = \left\lbrace 0;1\right\rbrace$$

$X_2(\omega_2) = \omega_2$ and $P(\omega_2) = 1/2 \quad \forall \, \omega_2 \in \left\lbrace 0; 1 \right\rbrace$.

$$ \omega_2 = \begin{cases}
0,&\text{halve the initial amount;}\\
1,&\text{double the initial amount.}
\end{cases}
$$

\end{experiment}

\begin{experiment}\label{xprmnt:three}

$X_3$ is a random variable on $(\Omega_3; \mathcal{F}_3;P_3)$. $X_3$ also represents the outcome of the toss of an unbiased coin. The sample space is again

$$\Omega_3 = \left\lbrace 0;1 \right\rbrace$$

$X_3(\omega_3) = \omega_3$ and $P(\omega_3) = 1/2 \quad \forall \, \omega_3 \in \left\lbrace 0; 1 \right\rbrace$.

$$ \omega_3 = \begin{cases}
0,&\text{allocate the first envelope to the agent;}\\
1,&\text{allocate the second envelope to the agent.}
\end{cases}
$$

\end{experiment}

\section{The agent}

The probability space that the agent would need to consider is

$$(\Omega; \mathcal{F}; P) = (\Omega_1 \times \Omega_2 \times \Omega_3; \mathcal{F}_1 \times \mathcal{F}_2 \times \mathcal{F}_3; P_1 \times P_2 \times P_3)$$ 

and the sample space for the agent (product measure space of the host) is expressed as

$$\Omega = \Omega_1 \times \Omega_2 \times \Omega_3.$$

Since the experiments are conducted independently, it can be said that $P = P_1 \times P_2 \times P_3$ and the proof of the existence of such a $P$ on $\mathcal{F}$ is an application of Carath\'eodory's extension theorem \citep[p24]{AL2010}.

\medskip

For any outcome $\left\lbrace c \right\rbrace = \left\lbrace x_1; \omega_2; \omega_3 \right\rbrace \in \mathcal{F}$, since the outcomes (events) $x_1$, $\omega_2$, and $\omega_3$ originate from each of the independent experiments, then

\begin{flalign}
&&P(\left\lbrace c \right\rbrace) &= P(\left\lbrace x_1;\omega_2;\omega_3 \right\rbrace)&&\notag\\
&& &= P(X_1 = x_1; X_2 = \omega_2; X_3 = \omega_3)&&\notag\\
&& &= P(X_1 = x_1) P(X_2 = \omega_2) P(X_3 = \omega_3).&&\label{qtn:event_probability}
\end{flalign}

%
%
%
   
\section{Envelope contents}

Knowledge of the process that the host adopted for the priming of the envelopes is crucial for the agent to determine an appropriate switching strategy. Most formulations of the problem centre on the \enquote{doubling-only} or \enquote{halving or doubling} processes. A \enquote{halving-only} process is introduced as the analog of the \enquote{doubling-only} process and it will assist with the derivation of the different switching strategies. 

\medskip

Having randomly identified an initial amount $X_1 = x_1$, the host assigns this to the first of the two indistinguishable envelopes. To the second envelope, the host will assign an amount $X_1^\prime$ that is dependent on the outcome of the toss of an unbiased coin. The transformation associated with the halving or doubling of the initial amount is as follows:

\begin{flalign}
&&X_1^\prime &= \left({1\over2}(1 - X_2) + 2X_2\right)X_1&&\label{qtn:trnsfrm_coin_1}
\end{flalign}

where

$$ X_2 = \begin{cases}
0,&\text{halve the initial amount;}\\
1,&\text{double the initial amount.}
\end{cases}
$$

Thus, the probability of any outcome $\left\lbrace c \right\rbrace=\left\lbrace x_1 ; \omega_2 ; \omega_3 \right\rbrace$ associated with the \enquote{halving or doubling} process after applying \ref{qtn:event_probability} is
  
\begin{flalign}
&&P(\left\lbrace c \right\rbrace) &= {1\over4}P(X_1 = x_1).&&\label{qtn:hod_probability}
\end{flalign}

The \enquote{halving or doubling} process is a combination of the \enquote{halving-only} and the \enquote{doubling-only} processes. Each of these processes can be realised by setting the relevant parameter $\omega_2$. 

\subsection{Doubling-only}

For the \enquote{doubling-only} process, the host only doubles the content of the first envelope and assigns this amount to the second envelope. Formally, $X_1^\prime = 2X_1$.

\medskip 

This can be realised by setting $X_2 (\omega_2) = \omega_2 = 1$. $\Omega_2 = \left\lbrace 1 \right\rbrace$ and $P(\omega_2) = 1$. Therefore, noting that $\left\lbrace c \right\rbrace = \left\lbrace x_1;1;\omega_3 \right\rbrace $, after applying (\ref{qtn:event_probability})

\begin{flalign}
&&P(\left\lbrace c \right\rbrace) &= {1\over2}P(X_1 = x_1).&&\label{qtn:do_probability}
\end{flalign}

\subsection{Halving-only}

For the \enquote{halving-only} process, the host only halves the content of the first envelope and assigns this amount to the second envelope. Formally, $X_1^\prime = X_1/2$.

\medskip 

This can be realised by setting $X_2(\omega_2) = \omega_2 = 0$. $\Omega_2 = \left\lbrace 0 \right\rbrace$ and $P(\omega_2) = 1$. Therefore, noting that $\left\lbrace c \right\rbrace =(x_1;0;\omega_3)$,  after applying (\ref{qtn:event_probability})

\begin{flalign}
&&P(\left\lbrace c \right\rbrace) &= {1\over2}P(X_1 = x_1).&&\label{qtn:do_probability}
\end{flalign}

\section{The envelope allocation}

After priming the envelopes, the host randomly selects one of the indistinguishable envelopes and allocates this envelope to the agent. The agent's envelope and its associated contents $y$ are represented by the random variable $Y$, while the complementary envelope and its associated contents $z$ are represented by the random variable $Z$. 

\medskip

The transformation process associated with the allocation of the envelopes and their respective contents can be expressed as follows:

\begin{flalign}
&&\begin{matrix}\textbf{C}\end{matrix}&=
\begin{bmatrix}
Y & Z
\end{bmatrix}&&\notag\\
&& &=
\begin{bmatrix}
X_1 & X_1^\prime
\end{bmatrix}
\begin{bmatrix}\textbf{A}
\end{bmatrix}&&\notag\\
&& &=\begin{matrix} X_1 \end{matrix}
\begin{bmatrix}
1 &  {1\over2}(1 - X_2) + 2X_2
\end{bmatrix}\begin{bmatrix}\textbf{A}
\end{bmatrix}&&\label{qtn:trnsfrm_coin_2_line_1}
\end{flalign}

where

$$\begin{matrix}\textbf{A}\end{matrix}=
\begin{bmatrix}
1-X_3 & X_3 \\
X_3 & 1- X_3 
\end{bmatrix},$$

$$X_2 = \begin{cases}
0,&\text{halve the initial amount;}\\
1,&\text{double the initial amount,}
\end{cases}$$

and

$$X_3 = \begin{cases}
0,&\text{allocate the first envelope to the agent;}\\
1,&\text{allocate the second envelope to the agent.}
\end{cases}$$

\section{The benefit of switching envelopes}

Since the objective of the agent is to identify an envelope-switching strategy that yields the greatest expected benefit for any observed $Y = y$, it is convenient to expand (\ref{qtn:trnsfrm_coin_2_line_1}) and accordingly derive the benefit function in the event of switching the randomly allocated envelope for the complementary envelope.

\medskip

The content of the agent's envelope is

\begin{flalign}
&&\begin{matrix}Y\end{matrix} &=\begin{matrix} X_1 \end{matrix}
\begin{bmatrix}
1 &  {1\over2}(1 - X_2) + 2X_2
\end{bmatrix}
\begin{bmatrix}
1-X_3 \\
X_3 
\end{bmatrix}.&&\label{qtn:agent}
\end{flalign}

The content of the complementary envelope is

\begin{flalign}
&&\begin{matrix}Z\end{matrix} &=\begin{matrix} X_1 \end{matrix}
\begin{bmatrix}
1 &  {1\over2}(1 - X_2) + 2X_2
\end{bmatrix}
\begin{bmatrix}
X_3 \\
1- X_3 
\end{bmatrix}.&&\label{qtn:complement}
\end{flalign}

The benefit of switching envelopes is

\begin{flalign}
&&B &= Z - Y&&\notag\\
&& &=\textbf{C}\begin{bmatrix}
-1 \\ 1
\end{bmatrix}&&\notag\\
&& &=\begin{bmatrix}
X_1 & X_1^\prime
\end{bmatrix}
\begin{bmatrix}
2X_3 - 1 \\ 1 - 2X_3
\end{bmatrix}&&\notag\\
&& &={1\over2} (2X_3 - 1) (1 - 3X_2)X_1.&&\label{qtn:benefit}
\end{flalign}

\medskip

It will be noted that this expression makes no direct reference to the content of the complementary envelope. The benefit is therefore entirely dependent on the initial amount identified by the host and the halving or doubling of this initial amount to determine the content of the second envelope. Finally, the random allocation of the envelopes, determines what is revealed in the allocated envelope.

%
%
%
\section{Summary}

\medskip

The host, having performed three experiments, assigns the initial amount to the first of the two envelopes and then assigned an amount to the second envelope, which is a function of the first amount. One of the primed envelopes is then randomly allocated to the agent. 

\medskip

The agent may receive the envelope containing $X_1 = x_1$ and refer to the content as \enquote{$Y$}; alternatively, he or she may receive the envelope containing the amount associated with $X_1^\prime$ and similarly refer to this as \enquote{$Y$}. 

\medskip

In all instances, the allocation process will commence with the identification of an initial amount $x_1$, represented by the random variable $X_1$, which can be assumed to originate from a density function $f_{X_1}(x_1)$. If the agent is not informed of the distribution function $F_{X_1}(x_1)$ from which the host sampled the initial amount, then the agent can assume or estimate the density function $f_{X_1}(x_1)$. Suffice to say that the agent will be informed and can in the presence of the density function $f_{X_1}(x_1)$, together with the content and allocation process derive an appropriate switching strategy.



\def\baselinestretch{1}

\chapter{The expected benefit of information}
\label{chptr:strategies}

\doublespacing

\begin{quotation}\small

Words are the source of misunderstandings.

\begin{flushright}
--- The Little Prince \citep[p65]{AS1943}
\end{flushright}

\end{quotation}

%
%
%

A switching strategy that optimises the expected outcome for any observed amount is possible if the agent is provided with information about the distribution function from which the initial amount was identified. This provision of information implies that the agent is informed about the content and allocation process as well as the distribution function from which the host randomly selected the initial amount. Not only is the distribution function essential for the agent to identify a strategy, but the content and allocation process is also crucial, as will be demonstrated in the strategy associated with the halving or doubling process. With this information, the agent can derive a game strategy that will enable him or her, for any observed amount, to identify the expected benefit and in so doing possess the information that will facilitate the decision to retain the allocated envelope and its contents or to switch to the complementary envelope and its contents.

\section{Essential information for the agent}

If the agent is not informed about the distribution function from which the original $X_1=x_1$ was selected, he or she may assume the existence of a prior density function. In addition, consequent to each play of the game, the agent may revise the guessed prior density function for the subsequent play of the game. However, irrespective of whether the agent has derived such a density function or not, the density function under consideration will be referred to as $f_{X_1}(x_1)$.

\medskip

Without loss of generality, any reference to distribution will imply a cumulative distribution function whether associated with a discrete -- or a continuous random variable. Similarly, there will be no notational distinction between a probability mass function associated with a discrete random variable and the alternative probability density function associated with a continuous random variable. The terms \enquote{discrete} and \enquote{continuous} are used to indicate the nature of the density function and similarly the distribution function.

\medskip

Since the objective of the game is to identify a strategy that will yield the greatest expected benefit, it is necessary to assume that all plays of the game are made by a risk-neutral and rational agent. 

\medskip

The agent will need to consider the content and allocation process adopted by the host and nature of the distribution function from which the host will sample the initial amount that will be assigned to the first envelope. The only information available to the host and the agent once the envelopes are allocated is the distribution from which the host sampled, the content and allocation process. The fact that the content of one envelope is double the content of the other, while true, is of no significance in determining the appropriate switching strategy. But, if one were informed that the agent had adopted the \enquote{doubling-only} process then it would be correct to say that the content of the second envelope is primed by doubling the amount assigned to the first envelope.   

\medskip

The agent will need to identify --- for the observed amount $y$ revealed in the randomly allocated envelope represented by the random variable $Y$ --- the expected benefit to be had by exercising the option of switching envelopes and therefore obtaining the expected content of the complementary envelope $Z$. 

\medskip

The benefit is represented by the random variable $B$ and is derived from the difference $B = Z - Y$ between the amount  (not known) in the complementary envelope, represented by the random variable $Z$, and the content (observed) of the randomly allocated envelope, represented by the random variable $Y$.

%
%
%

\section{The events associated with the sighted amount}

For the agent to observe an amount $Y = y$ the host would have had to have sampled an initial amount that may have been any one of the members of the set $x_1 \in \left\lbrace y/2;y;2y \right\rbrace$. It is the informed agent that can derive an appropriate switching strategy. This is immediately recognisable when the agent considers the {sub-$\sigma$-algebra} of relevant events induced by the observed amount $Y=y$ \citep{AR1972}. The conditional expected benefit of switching the allocated envelope for the complementary envelope is then easily derived when the appropriate event space is considered \citep{SA2013}. 

\medskip

$$ x_1 \in  \begin{cases}
\left\lbrace {y\over2};y \right\rbrace ,&\text{for \enquote{doubling-only};}\\
\left\lbrace y;2y \right\rbrace ,&\text{for \enquote{halving-only};}\\
\left\lbrace {y\over2};y;2y \right\rbrace ,&\text{for \enquote{halving or doubling}.}\\
\end{cases}
$$

\medskip

Suffice it to say, the contents of the envelopes are fixed and therefore the agent will reveal the content of the randomly allocated envelope to be $Y=y$. The agent will not know if this is the smaller amount and that the complementary envelope contains $Z=2y$ or the larger amount and that the complementary envelope then contains $Z=y/2$. While this statement may appear to relevant it will be revealed that this is nothing more than an interesting comment with no relevance to the agent when deriving the appropriate switching strategy. What is now relevant is the original amount identified by the host and how the content and allocation process impacts on this initial amount that ultimately determines what will be revealed in the envelope that is allocated to the agent.

\medskip

The following examples serve to explain the necessity of the agent, when observing $Y=y$ in the randomly allocated envelope, to consider that the host --- having adopted a halving or doubling process --- may have selected an initial amount $x_1 \in \left\lbrace y/2;y;2y \right\rbrace$. Each of these possibilities $x_1 \in \left\lbrace y/2;y;2y \right\rbrace$ is essential for the agent to identify a suitable switching strategy to realise the (greatest) expected benefit for any observed $Y = y$ in the randomly allocated envelope.

\medskip

Consideration will be given to the \enquote{doubling-only} strategy and, for the sake of expediency, a \enquote{halving-only} strategy will be introduced. By combining these two strategies a \enquote{halving or doubling} strategy will be derived. The \enquote{halving-only} strategy, as the complement of the \enquote{doubling-only} strategy, will be recognised as a distinct yet significant component of the \enquote{halving or doubling} strategy. To facilitate the derivation of the respective strategies for both discrete and continuous variables the concept of  a {sub-$\sigma$-algebra}, induced by $Y$, will be introduced for each of the strategies. 

\medskip

What is relevant to the agent, upon sighting the content of the allocated envelope, is not the event space $\mathcal{F}$ but the event space $\mathcal{F}^\prime$ induced by the observed amount $Y$ \citep[p250]{AR1972}. With this information the agent is able to determine the expected benefit of switching for any observed $Y=y$ \citep[p12]{NY1234}.

\medskip

%
%
%

 \subsection{A \enquote{doubling-only} strategy}

The agent observes $Y=y$ and considers the effects of the host having selected $x_1 \in \left\lbrace y/2;y\right\rbrace$. It is now essential for the agent to determine the values of $X_3 = \omega_3$ for which the allocated envelope would contain $Y=y$, when considering each of the initial assigned amounts $x_1 \in \left\lbrace y/2;y \right\rbrace$. $X_2 = \omega_2 = 1$. 

\begin{case}\label{case:four}

The host assigns the amount $X_1 = y/2$ to the first envelope . The host doubles the initial amount and assigns this amount $X_1^\prime = 2X_1 = y$ to the second envelope, and then allocates this second envelope to the agent, who will observe the amount $Y = y$. 

\end{case}

By substituting $X_1=y/2$, $Y=y$. and $\omega_2 = 1$ into (\ref{qtn:agent}) and solving for $\omega_3$, we find  

$$y = ((1-\omega_3) + 2\omega_3) {y\over2}.$$

The only possible solution to this equation is when $\omega_3 = 1$ (allocating the second envelope to the agent). The event $\left\lbrace c \right\rbrace = \left\lbrace y/2;1;1 \right\rbrace $ is thus realised; consequently, by substitution into (\ref{qtn:complement}), $Z=y/2$ and therefore

$$b(\left\lbrace c \right\rbrace ) = b(\left\lbrace y/2;1;1 \right\rbrace ) = -y/2$$ 
and 
$$P(\left\lbrace c \right\rbrace ) = P(\left\lbrace y/2;1;1 \right\rbrace ) = {1\over2}P(X_1 = y/2).$$
 
%
%
%
 
\begin{case}\label{case:five}

The host assigns the amount $X_1 = y$ to the first envelope .The host doubles the initial amount and assigns this amount $X_1^\prime = 2X_1 = 2y$ to the second envelope. If the first envelope is allocated, observe the amount $Y = y$. 

\end{case}

By substituting $X_1=y$, $Y=y$, and $\omega_2 = 1$ into (\ref{qtn:agent}) and solving for $\omega_3$, we find

$$y = ((1-\omega_3) + 2\omega_3)y.$$

The only possible solution to this equation is when $\omega_3 = 0$ (allocating the first envelope to the agent). The event $\left\lbrace c \right\rbrace = \left\lbrace y;1;0 \right\rbrace$ is thus realised; consequently, by substitution into (\ref{qtn:complement}), $Z={2y}$ and therefore

\bigskip

$$b(\left\lbrace c \right\rbrace ) = b(\left\lbrace y;1;0 \right\rbrace ) = y$$ 
and 
$$P(\left\lbrace c \right\rbrace ) = P(\left\lbrace y;1;0 \right\rbrace ) = {1\over2}P(X_1 = y).$$

\bigskip

Consequently, the set of events for consideration are

\begin{flalign}
&&C_y &= \left\lbrace \left\lbrace x_1;\omega_2;\omega_3 \right\rbrace :  \left\lbrace y/2;1;1 \right\rbrace ; \left\lbrace y;1;0 \right\rbrace \right\rbrace.&&\label{qtn:discrete_events_double}
\end{flalign}

%
%
%

 \subsection{A \enquote{halving-only} strategy}

The agent observes $Y=y$ and considers the effects of the host having selected $x_1 \in \left\lbrace y;2y\right\rbrace$. It is now essential for the agent to determine the values of $X_3 = \omega_3$ for which the allocated envelope would contain $Y=y$, when considering each of the initial assigned amounts $x_1 \in \left\lbrace y;2y \right\rbrace$. $X_2 = \omega_2 = 0$.

\begin{case}\label{case:seven}

The host assigns the amount $X_1 = y$ to the first envelope. The host halves the initial amount and assigns this amount $X_1^\prime = y/2$ to the second envelope. If the first envelope is allocated, then the agent will observe the amount $Y = y$. 

\end{case}

By substituting $X_1=y$, $Y=y$, and $\omega_2 = 0$ into (\ref{qtn:agent}) and solving for $\omega_3$, we find

$$y = (1-{1\over2}\omega_3)y.$$

The only possible solution to this equation is when $\omega_3 = 0$ (allocating the first envelope to the agent). The event $\left\lbrace c \right\rbrace = \left\lbrace y;0;0 \right\rbrace$ is thus realised; consequently, by substitution into (\ref{qtn:complement}), $Z={y/2}$ and therefore

 \bigskip

$$b(\left\lbrace c \right\rbrace) = b(\left\lbrace y;0;0 \right\rbrace ) = -y/2$$ 
and 
$$P(\left\lbrace c \right\rbrace ) = P(\left\lbrace y;0;0 \right\rbrace ) = {1\over2}P(X_1= y).$$

%
%
%

\begin{case}\label{case:six}

The host assigns the amount $X_1 = 2y$ to the first envelope. The host halves the initial amount and assigns this amount $X_1^\prime = y$ to the second envelope, and then allocates this second envelope to the agent, then the agent will observe the amount $Y = y$.

\end{case}

By substituting $X_1=2y$, $Y=y$. and $\omega_2 = 0$ into (\ref{qtn:agent}) and solving for $\omega_3$, we find  

$$y = (1-{1\over2}\omega_3) {2y}.$$

The only possible solution to this equation is when $\omega_3 = 1$ (allocating the second envelope to the agent). The event $\left\lbrace c \right\rbrace = \left\lbrace 2y;0;1 \right\rbrace $ is thus realised; consequently, by substitution into (\ref{qtn:complement}), $Z=2y$ and therefore
 
 \bigskip

$$b(\left\lbrace c \right\rbrace) = b(\left\lbrace 2y;0;1 \right\rbrace ) = y$$ 
and 
$$P(\left\lbrace c \right\rbrace) = P(\left\lbrace 2y;0;1 \right\rbrace ) = {1\over2}P(X_1 = 2y).$$

%
%
%

\medskip

Consequently, the set of events for consideration are

\begin{flalign}
&&C_y^\prime &= \left\lbrace \left\lbrace x_1;\omega_2;\omega_3 \right\rbrace:  \left\lbrace y;0;0 \right\rbrace ; \left\lbrace 2y;0;1\right\rbrace \right\rbrace.&&\label{qtn:discrete_events_halve}
\end{flalign}

%
%
%

\subsection{A \enquote{halving or doubling} strategy}

This strategy, it will be realised, is a combination of the \enquote{halving-only} and the \enquote{doubling-only} strategies.

\medskip

The agent observes $Y=y$ and considers the effects of the host having selected $x_1 \in \left\lbrace y/2;y;2y \right\rbrace$. It is now essential for the agent to identify the events that would describe the host's activities, namely determine the values of $\omega_2$ and $\omega_3$ for which the allocated envelope would contain $Y=y$, when he or she considers each of the initial assigned amounts $x_1 \in \left\lbrace y/2;y;2y \right\rbrace$.

\medskip

The combined events associated with the \enquote{halving-only} and \enquote{doubling-only} process are necessary for the \enquote{halving or doubling} process.

\medskip

\begin{flalign}
&&\mathcal{C}_y &= C_y \cup C_y^\prime&&\notag\\
&& &= \left\lbrace \left\lbrace x_1;\omega_2;\omega_3 \right\rbrace : \left\lbrace y/2;1;1 \right\rbrace ; \left\lbrace y;0;0 \right\rbrace ; \left\lbrace y;1;0 \right\rbrace ; \left\lbrace 2y;0;1\right\rbrace \right\rbrace.&&\label{qtn:discrete_events_either}
\end{flalign}
%
%
%

\subsection{Expected benefit associated with discrete variables}

The events of relevance to the agent for the identification of a strategy that will realise the expected benefit for any observed amount $Y=y$ are the members (events) of $\mathcal{C}_y$, $C_y^\prime$, or of $C_y$, depending on which content and allocation has been adopted by the host.

\bigskip

For the \enquote{doubling-only} strategy

\begin{flalign}
&&E(B \big\vert Y=y) &= E(B \big\vert C_y)&&\notag\\
&& &= {{\sum \limits_{\left\lbrace c \right\rbrace \in C_y} b(\left\lbrace c \right\rbrace) P(\left\lbrace c \right\rbrace)} \over {\sum \limits_{\left\lbrace c \right\rbrace \in C_y} P(\left\lbrace c \right\rbrace)}}&&\notag\\
&& &={{-{y\over2}P({X_1 = {y\over2}}) + yP(X_1 = y)}\over{P({X_1 = {y\over2}}) + P(X_1 = y)}}.&&\label{qtn:discrete_doubling}
\end{flalign}

For the \enquote{halving-only} strategy

\begin{flalign}
&&E(B \big\vert Y=y) &= E(B \big\vert C_y^\prime)&&\notag\\
&& &= {{\sum \limits_{ \left\lbrace c \right\rbrace \in C_y^\prime} b(\left\lbrace c \right\rbrace) P(\left\lbrace c \right\rbrace)} \over {\sum \limits_{\left\lbrace c \right\rbrace \in C_y^\prime} P(\left\lbrace c \right\rbrace)}}&&\notag\\
&& &={{-{y\over2}P({X_1 = y}) + yP(X_1 = 2y)}\over{P({X_1 = y}) + P(X_1 = 2y)}}.&&\label{qtn:discrete_halving}
\end{flalign}

For the \enquote{halving or doubling} strategy, from (\ref{qtn:discrete_doubling}) and (\ref{qtn:discrete_halving})

\begin{flalign}
&&E(B \big\vert Y=y) &= E(B \big\vert \mathcal{C}_y)&&\notag\\
&& &= {{\sum \limits_{\left\lbrace c \right\rbrace \in \mathcal{C}_y} b(\left\lbrace c \right\rbrace) P(\left\lbrace c \right\rbrace)} \over {\sum \limits_{\left\lbrace c \right\rbrace \in \mathcal{C}_y} P(\left\lbrace c \right\rbrace )}}&&\notag\\
&& &= {{\sum \limits_{\left\lbrace c \right\rbrace \in C_y \cup C_y^\prime} b(\left\lbrace c \right\rbrace) P(\left\lbrace c \right\rbrace)} \over {\sum \limits_{\left\lbrace c \right\rbrace \in C_y \cup C_y^\prime} P(\left\lbrace c \right\rbrace )}}&&\notag\\
&& &= {{\sum \limits_{\left\lbrace c \right\rbrace \in C_y} b(\left\lbrace c \right\rbrace) P(\left\lbrace c \right\rbrace) + \sum \limits_{ \left\lbrace c \right\rbrace \in C_y^\prime} b(\left\lbrace c \right\rbrace) P(\left\lbrace c \right\rbrace)} \over {\sum \limits_{\left\lbrace c \right\rbrace \in C_y} P(\left\lbrace c \right\rbrace) + \sum \limits_{\left\lbrace c \right\rbrace \in C_y^\prime} P(\left\lbrace c \right\rbrace)}}&&\notag\\
&& &={{-{y\over2}P({X_1 = {y\over2}}) + {y\over2}P(X_1 = y) + yP(X_1 =  2y)}\over{P({X_1 = {y\over2}}) + 2P(X_1 = y) + P(X_1 = 2y)}}.&&\label{qtn:discrete_either}
\end{flalign}

%
%
%
 
\section{Continuous random variables}

All discussions have centred on the agent observing an amount that is represented by a discrete random variable. It is possible that the observed amount can be represented by a continuous random variable. The probability of any observed amount represented by a continuous random variable, as is known, is zero \citep[p12]{NY1234}. However, if the observed amount is centred on an interval, the length of which tends to zero, then it is possible to derive a function for the continuous random variable analogous to that for the discrete random variable. This variant of the problem will permit the derivation of a strategy associated with an amount that is represented by a continuous random variable.

\medskip

Probabilities associated with continuous variables require a little more care in order to facilitate the derivation of appropriate switching strategies. Probabilities associated with intervals are easily dealt with for continuous variables. This feature will be exploited by considering intervals of lengths, that are proportionately $\varepsilon$, $2\varepsilon$, and $4\varepsilon$ in length. For any chosen interval, the limit as $\varepsilon$ tends to zero will be exploited. As noted by \citep[p241]{AR1972}, \citep[p61]{MA1974}, \citep[p30]{BK1995}, and \citep[p10]{BJ1995}, the approximation does yield the appropriate switching function. 

\medskip

Where $f_{X_1}(x)$ is continuous, the following approach is adopted. For the agent to observe an amount in the allocated envelope, from the interval $y-\varepsilon < Y \le y + \varepsilon$,  the agent would have to consider the activities of the host in drawing, depending on the content and allocation process,  an initial amount from the set of intervals:

\subsection{Probability}\label{ppndx:prbblty}

$P(X=x)$ is not true for a continuous probability density function $f_X (x)$. However, it can be approximated as described by \citet[241--261]{AR1972} and \citet[61]{MA1974}.

\bigskip

%
%

\begin{tikzpicture}[node distance=1cm, auto,]

\draw (1,0) plot[smooth] coordinates {(3,1.73205081) (4,2) (5,2.23606798) (6,2.44948947) (7,2.645575131) (8,2.82842712) (9,3) (10,3.16227766) (11,3.31662479) (12,3.46410162) (13,3.60555128)};

\draw (4,2.5) -- (4,0) node[below] {$x-\varepsilon$};
\draw (8,2.82842712) node[above] {$f_{X}(x)$} -- (8,0) node[below] {$x$};
\draw (12,3.96410162) -- (12,0) node[below] {$x+\varepsilon$};

\draw (3,0.25) -- (13,0.25);

\end{tikzpicture}

\bigskip

\begin{flalign*}
&&f_{X}(x)=& {dF_{X}(x)}\over{dx}&\\
&&=& \lim_{\varepsilon \to 0} {{F_{X}(x + \varepsilon) - F_{X}(x - \varepsilon)} \over {2\varepsilon}}&
\end{flalign*}

Hence,

\begin{flalign*}
&&f_{X}(x) 2\varepsilon \approx {F_{X}(x + \varepsilon) - F_{X}(x - \varepsilon)} =& P({x - \varepsilon} < X \leq {x + \varepsilon})&\\
&&=& P(X \le x + \varepsilon) - P(X \le x - \varepsilon)&\\
&& &= \int_{y - \varepsilon}^{y + \varepsilon} f_{X_1}(x_1) dx_1&&\notag
\end{flalign*}


\subsection{Expectation}\label{ppndx:xpcttn}

The {$\sigma$-algebra} induced by the observed amount is very convenient. For the random object $Y: (\Omega; \mathcal{F}) \longrightarrow (\Omega^\prime; \mathcal{F}^\prime)$, then the {$\sigma$-algebra} induced by $Y$ is given by $\mathcal{F}(Y) = Y^{-1} (\mathcal{F}^\prime)$  \citep[p250]{AR1972}. And, the conditional expectation $E(B\big\vert Y=y)$ is well defined for any number $Y=y$ even though $Y=y$ may be a probability zero event \citep[p12]{NY1234}.

It would be beneficial to derive an appropriate definition for the expectation of continuous variable $X$ over a stated interval $y-\varepsilon < X \le y +\varepsilon$ \citep[265]{RA1970}. Then 

\begin{flalign*}
&& E(X=y) &\approx \lim_{\varepsilon \to 0}E(X \big\vert y - \varepsilon < X \le y + \varepsilon)&&\\
&& &=\lim_{\varepsilon \to 0} {{\int_{y-\varepsilon}^{y + \varepsilon} x f_X(x) dx} \over {\int_{y-\varepsilon}^{y + \varepsilon} f_X(x) dx}}&&\\
&& &=\lim_{\varepsilon \to 0} {{yf_X(y)2\varepsilon} \over {f_X(y)2\varepsilon}}&&\\
&& &=y&&
\end{flalign*}

%
%
%

 \subsection{A \enquote{doubling-only} strategy}
 
%
%
%

For the \enquote{doubling-only} process

$$C_y = \mathop{\bigcup}_{\omega_3 \in \left\lbrace 0;1 \right\rbrace} \left\lbrace \left\lbrace x_1,1,\omega_3 \right\rbrace : 2^{-\omega_3}(y-\varepsilon) < x_1 \le 2^{-\omega_3}(y + \varepsilon) \right\rbrace.$$

\bigskip

Noting, from (\ref{qtn:benefit}), that

$$ b(\left\lbrace c \right\rbrace ) =b(\left\lbrace x_1;1;\omega_3 \right\rbrace)= \begin{cases}
x_1,&\text{for $\omega_3 = 0$;}\\
-x_1,&\text{for $\omega_3 = 1$,}
\end{cases}
$$

and following from (\ref{qtn:event_probability}) then 

\begin{flalign}
&&E(B \big\vert Y=y) &\approx \lim_{\varepsilon \to 0} E(B \big\vert y - \varepsilon < Y \le y + \varepsilon)&&\notag\\
&& &= \lim_{\varepsilon \to 0}  E(B \big\vert C_y)&&\notag\\
&& &= \lim_{\varepsilon \to 0}  {{\int_{C_y} b(\left\lbrace c \right\rbrace ) f_{X_1}(x_1) dx_1} \over {\int_{C_y} f_{X_1}(x_1) dx_1}}&&\notag\\
&& &= \lim_{\varepsilon \to 0} {{ -\int_{{y \over 2}-{\varepsilon \over 2}}^{{y \over 2}+{\varepsilon \over 2}} {x_1} f_{X_1}(x_1) dx_1 + \int_{y-\varepsilon}^{y+\varepsilon} {x_1}  f_{X_1}
(x_1) dx_1} \over {\int_{{y \over 2}-{\varepsilon \over 2}}^{{y \over 2} +{\varepsilon \over 2}} f_{X_1}(x_1) dx_1 + \int_{y-\varepsilon}^{y+\varepsilon} f_{X_1}(x_1) dx_1}}&&\notag\\
&& &\approx \lim_{\varepsilon \to 0}  {{-{y\over2}f_{X_1}({y\over2})\varepsilon + yf_{X_1}(y)2\varepsilon} \over {f_{X_1}({y\over2})\varepsilon + f_{X_1}(y)2\varepsilon}}&&\notag\\
&& &= {{-{y\over2}f_{X_1}({y\over2}) + 2yf_{X_1}(y)} \over {f_{X_1}({y\over2}) + 2f_{X_1}(y)}}.&&\label{qtn:continuous_doubling}
\end{flalign}

It is worth noting that this function is in agreement with the function derived by \citet[p10]{BJ1995} for the $E(Z \big\vert Y = y)$. By subtracting the observed amount $Y = y$ from the expression derived by \citet{BJ1995}, the function as derived above can be obtained.  

%
%
%

\subsection{A \enquote{halving-only} strategy}

For the \enquote{halving-only} process

$$C_y^\prime = \mathop{\bigcup}_{\omega_3 \in \left\lbrace 0;1 \right\rbrace} \left\lbrace \left\lbrace x_1,0,\omega_3 \right\rbrace : 2^{\omega_3} (y-\varepsilon) < x_1 \le 2^{\omega_3}(y + \varepsilon) \right\rbrace.$$

\bigskip

Noting, from (\ref{qtn:benefit}), that

$$ b(\left\lbrace c \right\rbrace ) =b( \left\lbrace x_1;0;\omega_3 \right\rbrace )= \begin{cases}
-{1\over2} x_1,&\text{for $\omega_3 = 0$;}\\
{1\over2} x_1,&\text{for $\omega_3 = 1$,}
\end{cases}
$$

and following from (\ref{qtn:event_probability}) then 

\begin{flalign}
&&E(B \big\vert Y=y) &\approx \lim_{\varepsilon \to 0} E(B \big\vert y - \varepsilon < Y \le y + \varepsilon)&&\notag\\
&& &= \lim_{\varepsilon \to 0}  E(B \big\vert C_y^\prime)&&\notag\\
&& &= \lim_{\varepsilon \to 0}  {{ \int_{C_y^\prime} b(\left\lbrace c \right\rbrace) f_{X_1}(x_1) dx_1} \over {\int_{C_y^\prime} f_{X_1}(x_1) dx_1}}&&\notag\\
&& &=\lim_{\varepsilon \to 0}{{ -{1\over2}\int_{y - \varepsilon}^{y + \varepsilon} {x_1} f_{X_1}(x_1) dx_1 + {1\over2}\int_{2y-2\varepsilon}^{2y+2\varepsilon} {x_1}  f_{X_1}(x_1) dx_1} \over {\int_{y - \varepsilon}^{y + \varepsilon} f_{X_1}(x_1) dx_1 + \int_{2y-2\varepsilon}^{2y+2\varepsilon} f_{X_1}(x_1) dx_1}}&&\notag\\
&& &\approx \lim_{\varepsilon \to 0}  {{-yf_{X_1}(y)\varepsilon + 4yf_{X_1}(y)\varepsilon} \over {2f_{X_1}(y)\varepsilon + 4f_{X_1}(2y)\varepsilon}}&&\notag\\
&& &= {{-yf_{X_1}(y) + 4yf_{X_1}(2y)} \over {2f_{X_1}(y) + 4f_{X_1}(2y)}}.&&\label{qtn:continuous_halving}
\end{flalign}

%
%

\subsection{A \enquote{halving or doubling} strategy}

The combined events associated with the \enquote{halving-only} and \enquote{doubling-only} process are necessary for the \enquote{halving or doubling} process.

\medskip

$$\mathcal{C}_y = C_y \cup C_y^\prime.$$

However, if $f_{X_1}(x)$ were continuous, the following approach would be adopted. For the agent to observe an amount in the allocated envelope, from the interval $y-\varepsilon < Y \le y + \varepsilon$, he or she would, assuming the \enquote{halving or doubling} process, have to consider the activities of the host in drawing an initial amount from the set of intervals:

\begin{flalign}
&&E(B \big\vert Y=y) &\approx \lim_{\varepsilon \to 0} E(B \big\vert y - \varepsilon < Y \le y + \varepsilon)&&\notag\\
&& &= \lim_{\varepsilon \to 0} E(B \big\vert \mathcal{C}_y)&&\notag\\
&& &= \lim_{\varepsilon \to 0}  E(B \big\vert C_y \cup C_y^\prime)&&\notag\\
&& &= \lim_{\varepsilon \to 0} {{\int_{C_y} b(\left\lbrace c \right\rbrace) f_{X_1}(x_1) dx_1 + \int_{C_y^\prime} b(\left\lbrace c \right\rbrace) f_{X_1}(x_1) dx_1} \over {\int_{C_y} f_{X_1}(x_1) dx_1 + \int_{C_y^\prime} f_{X_1}(x_1) dx_1}}&&\notag\\
&& &=\lim_{\varepsilon \to 0}{{ -\int_{{y \over 2}-{\varepsilon \over 2}}^{{y \over 2}+{\varepsilon \over 2}} {x_1} f_{X_1}(x_1) dx_1 + {1\over2} \int_{y-\varepsilon}^{y+\varepsilon} {x_1} f_{X_1}(x_1) dx_1 + {1\over2} \int_{2y-2\varepsilon}^{2y+2\varepsilon} {x_1} f_{X_1}(x_1) dx_1} \over {\int_{{y \over 2}-{\varepsilon \over 2}}^{{y \over 2} +{\varepsilon \over 2}} f_{X_1}(x_1) dx_1 + 2\int_{y-\varepsilon}^{y+\varepsilon} f_{X_1}(x_1) dx_1 + \int_{2y-2\varepsilon}^{2y+2\varepsilon} f_{X_1}(x_1) dx_1 }}&&\notag\\
&& &\approx \lim_{\varepsilon \to 0}  {{-{y\over2}f_{X_1}({y\over2})\varepsilon + {y\over2}f_{X_1}(y)2\varepsilon + yf_{X_1}(2y)4\varepsilon} \over {f_{X_1}({y\over2})\varepsilon + 2f_{X_1}(y)2\varepsilon + f_{X_1}(2y)4\varepsilon}}&&\notag\\
&& &= {{-{y\over2}f_{X_1}({y\over2}) + yf_{X_1}(y) + 4yf_{X_1}(2y)} \over {f_{X_1}({y\over2}) + 4f_{X_1}(y) + 4f_{X_1}(2y)}}.&&\label{qtn:continuous_either}
\end{flalign}

%
%
%

Hence, the identification of the content assigned to the envelopes and the subsequent allocation of the primed envelopes is independent of any game strategy identified by the agent. However, the benefit of switching must be conditional upon the realised (observed) amount $Y = y$, and not the distinctly different $Y = X_1 = x_1$ or $Y=X_1^\prime = (1/2(1-\omega_2) + 2\omega_2)X_1$. The only events that generate $Y=y$, the amount observed by the agent in the allocated envelope, are associated with the host selecting an initial amount $X_1=x_1$ where $x_1 \in \left\lbrace y/2;y;2y \right\rbrace$ and assigning this amount to the first envelope.

%
%
%
%

\section{A switching strategy}

If the host were to impose bounds on the amounts allocated, then the agent would need to consider what impact this would have on the host's activities and consequently on his or her own strategy.

\medskip

The agent, if informed about the density function from which the monetary amounts were drawn should play as follows. If a lower bound $x_l$ is specified and all monetary amounts are greater than or equal to this lower bound $x_l$, then if the content of the chosen envelope is revealed to be less than $2x_l$, the agent should switch. Similarly, if an upper bound $x_u$ is specified and all monetary amounts are less than or equal to this upper bound $x_u$, then if the content is revealed to be greater than ${x_u}/2$, the agent should not switch. For all other observed amounts $2x_l < y \le {x_u}/2$, the agent would need to consider $E(B \big\vert Y = y)$. If $E(B \big\vert Y = y) < 0$, then the agent should not switch, whereas if $E(B \big\vert Y = y) > 0$, then the agent should switch, otherwise he or she should be indifferent.

\medskip 

The switching strategy $s(y)$ can be expressed succinctly using an indicator function:

\begin{flalign}
&&s(y) &= yI_{[{x_l};{2x_l}]} + E(B \big\vert Y = y)I_{[2{x_l};{1\over2}{x_u}]} - {y\over 2}I_{[{1\over 2}{x_u};{x_u}]}.&&\label{qtn:indicator}
\end{flalign}

\medskip

The strategy for any observed ${x_l} \le y \le {x_u}$ is as follows: For $s(y) < 0$, the agent should not switch. For $s(y) > 0$, the agent should switch, and for $s(y) = 0$, the agent should be indifferent.

\medskip

It will be necessary for the agent to consider what the impact will be at the boundaries $2x_l$ and ${x_u}/2$. At these transition points it may necessitate that the agent evaluate the expected benefit of switching, before deciding a particular action.


\section{Other strategies}

There are two other possible strategies that can be considered. One is associated with the host priming the first envelope and allocating the envelope to the agent before assigning an amount to the second and complementary envelope. And reciprocal to this is when the host primes the first envelope and retains the envelope as the complement. The second envelope is then primed before allocating it to the agent. It will be noted that the assigning of the amount to the second envelope, in both these instances, must be associated with the \enquote{halving or doubling} process.

\subsection{Allocate the first envelope}

A strategy associated with the former is nothing more than a simple lottery in which you have for certain the amount $Y=y$ and the complementary envelope contains the expected amount $Z = y/4 + y$. In this instance there is an obvious benefit $b(y) = y/4$ to be had for switching. The events of relevance in this instance are $C_y = \left\lbrace \left\lbrace x_1;\omega_2;0 \right\rbrace : \left\lbrace y;0;0 \right\rbrace ; \left\lbrace y;1;0 \right\rbrace \right\rbrace$. Here the content and process excludes switching $\omega_3 = 0$ and the probability of the initial amount is of no relevance in the switching strategy. The benefit 

\begin{flalign}
&&b(\left\lbrace c \right\rbrace ) &= b(\left\lbrace x_1;\omega_2;0 \right\rbrace )&&\notag\\
&& &= \begin{cases}
-{1\over2}x_1,&\text{if $\omega_2 = 0$;}\\
x_1,&\text{if $\omega_2 = 1$.}
\end{cases}&&\notag
\end{flalign}

 For discrete variables the expected benefit would be

\begin{flalign}
&&E(B \big\vert Y=y) &= E(B \big\vert C_y)&&\notag\\
&& &={{\sum \limits_{\left\lbrace c \right\rbrace  \in C_y} b(\left\lbrace c \right\rbrace ) P(\left\lbrace c \right\rbrace)} \over {\sum \limits_{\left\lbrace c \right\rbrace \in C_y} P(\left\lbrace c \right\rbrace )}}&&\notag\\
&& &={{-{y\over2}P(X_1 = y) +  yP(X_1 = y)}\over{P(X_1 = y) + P(X_1 = y)}}&&\notag\\
&& &={y\over4} > 0.&&\label{qtn:llct_first}
\end{flalign}

While for continuous variables

\begin{flalign}
&&C_y &= \left\lbrace \left\lbrace x_1;\omega_2;0 \right\rbrace :  y-\varepsilon < x_1\le y + \varepsilon \right\rbrace &&\notag
\end{flalign}

and following from (\ref{qtn:event_probability}) then 

\begin{flalign}
&&E(B \big\vert Y=y) &\approx \lim_{\varepsilon \to 0} E(B \big\vert y - \varepsilon < Y \le y + \varepsilon)&&\notag\\
&& &= \lim_{\varepsilon \to 0}  E(B \big\vert C_y)&&\notag\\
&& &= \lim_{\varepsilon \to 0}  {{\int_{C_y} b(\left\lbrace c \right\rbrace ) f_{X_1}(x_1) dx_1} \over {\int_{C_y} f_{X_1}(x_1) dx_1}}&&\notag\\
&& &= \lim_{\varepsilon \to 0} {{ -{1\over2}\int_{y-\varepsilon}^{y + \varepsilon} {x_1} f_{X_1}(x_1) dx_1 + \int_{y-\varepsilon}^{y+\varepsilon} {x_1}  f_{X_1}
(x_1) dx_1} \over {\int_{y - \varepsilon}^{y + \varepsilon} f_{X_1}(x_1) dx_1 + \int_{y-\varepsilon}^{y+\varepsilon} f_{X_1}(x_1) dx_1}}&&\notag\\
&& &= \lim_{\varepsilon \to 0} {{{1\over2}\int_{y-\varepsilon}^{y + \varepsilon} {x_1} f_{X_1}(x_1) dx_1} \over {2\int_{y - \varepsilon}^{y + \varepsilon} f_{X_1}(x_1) dx_1}}&&\notag\\
&& &\approx \lim_{\varepsilon \to 0}  {{{1\over2}yf_{X_1}(y)2\varepsilon} \over {2f_{X_1}(y)2\varepsilon}}&&\notag\\
&& &= {y\over4} > 0.&&\label{qtn:llct_continuous_first}
\end{flalign}

and hence, irrespective of the type of variable, the same expected benefit is realised and it would therefore always be beneficial for the agent to switch envelopes.

\medskip

This strategy is alluded to by \citet[p212]{CJ1992} and amounts considering a gamble on the outcome of a fair coin at payoff of double or half, while in receipt of a certain amount $Y=y$.

\subsection{Allocate the second envelope}

A strategy associated with the later is obtained similarly. The agent will reveal the content of the allocated envelope to be $Y=y$. In this instance for the host may have generated an initial amount $X_1 = y/2$ or $X_1 = 2y$. The only possible host activity is associated with the events $C_y = \left\lbrace \left\lbrace x_1;\omega_2;1 \right\rbrace : \left\lbrace y/2;1;1 \right\rbrace; \left\lbrace 2y;0;1 \right\rbrace \right\rbrace$. Here the process includes switching $\omega_3 = 1$ and the probability of the initial amounts $X_1 = y/2$ and $X_1 = 2y$ is significant in determining a switching strategy. Any strategy associated with such a content and allocation process is dependent on the distribution function. The benefit would be

\begin{flalign}
&&b(\left\lbrace c \right\rbrace ) &= b(\left\lbrace x_1;\omega_2;1 \right\rbrace )&&\notag\\
&& &= \begin{cases}
{1\over2}x_1,&\text{if $\omega_2 = 0$;}\\
-x_1,&\text{if $\omega_2 = 1$.}
\end{cases}&&\notag
\end{flalign}

For discrete variables the expected benefit would be

\begin{flalign}
&&E(B \big\vert Y=y) &= E(B \big\vert C_y)&&\notag\\
&& &={{\sum \limits_{\left\lbrace c \right\rbrace \in C_y} b(c) P(c)} \over {\sum \limits_{\left\lbrace c \right\rbrace \in C_y} P(c)}}&&\notag\\
&& &={{-{y\over2}P(X_1 = {y\over2}) + yP(X_1 = 2y)}\over{P(X_ 1= {y\over2}) + P(X_1 = 2y)}}.&&\label{qtn:llct_second}
\end{flalign}

Switching in this instance is now conditional upon the probability density function and it will be always be beneficial to switch when 

$$2P(2y) > P({y\over2}).$$ 

While for continuous variables

$$C_y = \left\lbrace \left\lbrace x_1;\omega_2;1 \right\rbrace :  2^{1-2\omega_2}(y-\varepsilon) < x_1\le 2^{1-2\omega_2}(y + \varepsilon) \right\rbrace$$

then

\begin{flalign}
&&E(B \big\vert Y=y) &\approx \lim_{\varepsilon \to 0} E(B \big\vert y - \varepsilon < Y \le y + \varepsilon)&&\notag\\
&& &= \lim_{\varepsilon \to 0}  E(B \big\vert C_y)&&\notag\\
&& &= \lim_{\varepsilon \to 0}  {{ \int_{C_y} b(\left\lbrace c \right\rbrace ) f_{X_1}(x_1) dx_1} \over {\int_{C_y} f_{X_1}(x_1) dx_1}}&&\notag\\
&& &=\lim_{\varepsilon \to 0}{{{1\over2}\int_{2y - 2\varepsilon}^{2y + 2\varepsilon} {x_1} f_{X_1}(x_1) dx_1 - \int_{{y\over2}-{\varepsilon\over2}}^{{y\over2}+{\varepsilon\over2}} {x_1}  f_{X_1}(x_1) dx_1} \over {\int_{2y - 2\varepsilon}^{2y + 2\varepsilon} f_{X_1}(x_1) dx_1 + \int_{{y\over2}-{\varepsilon\over2}}^{{y\over2}+{\varepsilon\over2}} f_{X_1}(x_1) dx_1}}&&\notag\\
&& &\approx \lim_{\varepsilon \to 0}  {{-{y\over2}f_{X_1}({y\over2})\varepsilon + 4yf_{X_1}(2y)\varepsilon} \over {f_{X_1}({y\over2})\varepsilon + 4f_{X_1}(2y)\varepsilon}}&&\notag\\
&& &= {{-{y\over2}f_{X_1}({y\over2}) + 4yf_{X_1}(2y)} \over {f_{X_1}({y\over2}) + 4f_{X_1}(2y)}}.&&\label{qtn:llct_continuous_second}
\end{flalign}

Switching in this instance is now conditional upon the probability density function and it will be always be beneficial to switch when 

$$8f_{X_1}(2y) > f_{X_1}({y\over2}).$$ 

\section{Summary}

It is essential for the agent to be informed about the content and allocation process adopted by the host in order to derive an appropriate switching strategy. As we have demonstrated, the envelope-switching strategy is dependent on the content and allocation process adopted by the host, and the distribution function from which the host sampled the initial amount assigned to the first envelope. Once the agent is in possession of this information, the strategy is easily derived. This information includes not only the sighting of the content of the allocated envelope, but also the adopted content and allocation process.

\medskip

A switching strategy to optimise the expected outcome for any observed $Y=y$ is possible if the agent is provided with this information.
Moreover, the agent also needs to be informed about the bounds $x_l$ and $x_u$ if they are applied. Hence, two distinct strategies are available depending on the content and allocation process. Each strategy, depending on the distribution function from which the initial amount is chosen, may result in a different decision for any observed amount in the allocated envelope. 

\medskip

From the derived strategies it will be realised that the content of the complementary envelope has no impact on the switching strategy. The content of the complementary envelope is of nothing more than curiosity value?


\def\baselinestretch{1}

\chapter{{Envelope-switching} strategies}
\label{chptr:examples}

Some distribution functions are labelled paradoxical by \citet{BJ1995}. Such distributions are by their nature not paradoxical. Such an attribute is only possible because the rate of monetary benefit exceeds the probability of such an event. As demonstrated by the Weibull density function, some distribution functions that are not paradoxical. The choice of density functions to demonstrate the aspects of the two-envelope problem for discrete and continuous random variables is now presented. Many of these have already been explored in the literature and are repeated here.

%
%
%
%

\section{Uniform distribution}

Suppose the agent is informed that the host has chosen to randomly sample the initial amount from the standard uniform probability density function and that $$f_{X_1}(x_1) = 1 \quad \forall \; 0 < x_1 \le 1.$$

Since $X_1$ is a continuous random variable, applying (\ref{qtn:continuous_either}) and (\ref{qtn:hod_probability}) for the halving or doubling process, we find

$$E(B \big\vert Y = y) = {{-{y\over2} + y + 4y}\over{1 + 4 + 4}} = {y\over2}.$$

Meanwhile, by applying \ref{qtn:continuous_doubling} and (\ref{qtn:do_probability}) to the doubling-only process, we find

$$E(B \big\vert Y = y) = {{-{y\over2} + 2y}\over{1 + 2}} = {y\over2}.$$

Therefore, for either of the content and allocation processes, the indicator function (\ref{qtn:indicator}) yields the same envelope-switching strategy:

$$s(y) =\begin{cases}
{y\over2},&\text{if $0 < y \le {1\over2}$, therefore switch;}\\
-{y\over2},&\text{if ${1\over2} < y \le 1$, therefore do not switch.}
\end{cases}$$

%
%
%
%

\section{Weibull distribution}

Suppose the agent is informed that the host has chosen to randomly sample the initial amount from a Weibull probability density function and that $$X \sim f_X(x,\lambda,k) = {k \over \lambda}{\left ( x \over \lambda \right ) ^{k-1}} {e^{-{\left (x \over \lambda \right ) ^k}}} \quad \forall \, x \ge 0.$$

We set the parameter $$\lambda = {1\over 2}$$ and for $k = 2$ the distribution function is also referred to as the \enquote{Rayleigh distribution}. 

Then $$X \sim f_X(x) = f_X(x,{1\over2},2) =   8x {e^{-{\left (2x \right ) ^2}}} \quad \forall \, x > 0.$$

The expected benefit for an observed $Y = y$ is then expressed as follows.

\subsection{Strategy associated with the \enquote{doubling-only} process}

Applying (\ref{qtn:continuous_doubling})

 \begin{flalign*}
&&E(B \big\vert Y=y) &= {{-{y\over2}f_{X_1}({y\over2}) + 2yf_{X_1}(y)} \over {f_{X_1}({y\over2}) + 2f_{X_1}(y)}}&&\\
&& &=y{{8e^{-3y^2} - 1}\over{8e^{-3y^2} + 2}}.&&\\
\end{flalign*}

The concept of an exchange condition $e(y)$ has been introduced by \citet[p175]{BK2001} and as a result 

$$E(B\big\vert Y=y) = {{e(y)} \over {8e^{-3y^2} + 2}}.$$ 

The denominator in $E(B\big\vert Y=y)$ will always be positive. Solving for $y$ in the exchange condition $e(y) = 0$. Then $8e^{-3y^2} - 1 = 0$ and therefore $y = \sqrt{\ln 2}.$

Consequently the switching strategy $s(y)$, when observing $Y=y$, is
  
$$s(y) =\begin{cases}
\text{Switch},&\text{if $y < \sqrt{\ln 2}$;}\\
\text{Indifferent},&\text{if $y = \sqrt{\ln 2}$;}\\
\text{Do not switch},&\text{if $y > \sqrt{\ln 2}.$}
\end{cases}$$  

\bigskip

\subsection{Strategy associated with the \enquote{halving and doubling} process}

Applying (\ref{qtn:continuous_either})

 \begin{flalign*}
&&E(B \big\vert Y=y) &= {{-{y\over2}f_{X_1}({y\over2}) + yf_{X_1}(y) + 4yf_{X_1}(2y)} \over {f_{X_1}({y\over2}) + 4f_{X_1}(y) + 4f_{X_1}(2y)}}&&\\
&& &=y{{32e^{-15y^2} + 4e^{-3y^2} - 1} \over {32e^{-15y^2} + 16e^{-3y^2} + 2}}
\end{flalign*}

Solving for $y$ in the following exchange condition

$$e(y) = 32e^{-15y^2} + 4e^{-3y^2} - 1 = 0.$$ 

Then $$y \approx 0.686511 = k.$$

Consequently, for the \enquote{halving or doubling} process, the switching strategy $s(y)$, when observing $Y=y$, is
  
$$s(y) =\begin{cases}
\text{Switch},&\text{if $y < k$;}\\
\text{Indifferent},&\text{if $y = k$;}\\
\text{Do not switch},&\text{if $y > k$}
\end{cases}$$ 

\bigskip 

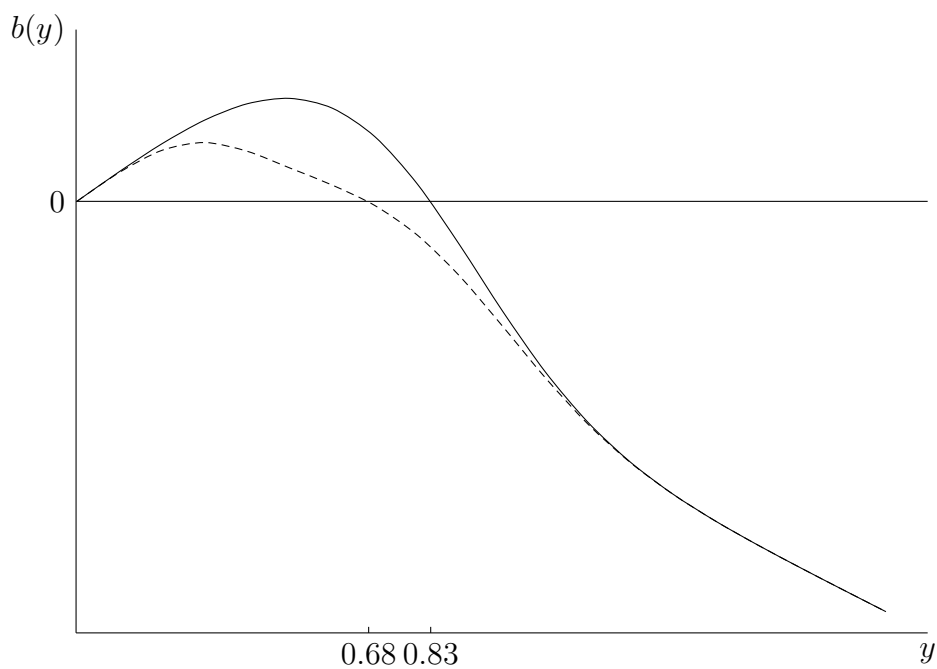
\begin{figure}[!htb]  

\bigskip

\centering

\begin{tikzpicture}


\draw [scale=0.8] [densely dashed] plot[smooth] coordinates{
(0.014, 7.152857003)
(0.714, 7.6339326256)
(1.414, 7.9993475136)
(2.114, 8.1277763315)
(2.814, 7.9829993994)
(3.514, 7.7059180358)
(4.214, 7.4214035388)
(4.914, 7.0824436216)
(5.614, 6.5904193166)
(6.314, 5.9060038644)
(7.014, 5.0802179357)
(7.714, 4.2290898753)
(8.414, 3.4612272741)
(9.114, 2.8217136751)
(9.814, 2.2972890631)
(10.514, 1.8519223737)
(11.214, 1.4524305038)
(11.914, 1.0765375406)
(12.614, 0.7116806584)
(13.314, 0.351577067)};\label{do}


\draw [scale=0.8] plot[smooth] coordinates{
(0.014, 7.1528571017)
(0.714, 7.6473483669)
(1.414, 8.1089085308)
(2.114, 8.4995776302)
(2.814, 8.770615585)
(3.514, 8.8598055245)
(4.214, 8.6965643848)
(4.914, 8.2233368936)
(5.614, 7.435725943)
(6.314, 6.4182792629)
(7.014, 5.3294287389)
(7.714, 4.3259219673)
(8.414, 3.4916819386)
(9.114, 2.8296532869)
(9.814, 2.2990459047)
(10.514, 1.8522582369)
(11.214, 1.4524866004)
(11.914, 1.0765457776)
(12.614, 0.7116817252)
(13.314, 0.351577189)};\label{hod}

\draw [scale=0.8] (0,0) -- (0,10) node[left]{$b(y)$};
\draw [scale=0.8] (0,0) -- (14,0) node[below]{$y$};

\draw [scale=0.8] (0, 7.1528571017) node[left] {$0$} -- (14,7.1528571017);

\draw [scale=0.8] (4.81,0) node[below]{$0.68$} -- (4.81, 0.1);
\draw [scale=0.8] (5.83,0) node[below]{$0.83$} -- (5.83, 0.1);

\end{tikzpicture}

\caption[The expected benefit $b(y) = E(B \big\vert Y=y)$ for the agent observing $Y=y$ \ref{do} and \ref{hod}]{Weibull: The expected benefit $b(0.686511) \approx 0$ when the 
\enquote{halving or doubling} strategy is adopted and $b(\sqrt{\ln 2}) = 0$ when the \enquote{doubling-only} strategy is adopted.}\label{fig:weibull}

\end{figure}
 
In this example, it is noted that the benefit functions associated with the different content and allocation processes highlight the difference in strategies. However, both in Figure~\ref{fig:weibull} indicate that the expected benefit becomes negative as the observed value increases and that the agent should therefore not switch.
  
%
%
%

\section{Broome's approach}

The following examples, that are described by \citet[p7]{BJ1995}, are associated with a \enquote{doubling-only} content and allocation process.

\subsection{Discrete variable}

$$f_{X_1}(x_1) = {{2^n} \over {3^{n+1}}} \quad \forall \; x_1 = 2^n, n \in \mathbb{N}_0.$$

Applying (\ref{qtn:discrete_doubling})

\begin{flalign*}
&&E(B \big\vert Y=y) &= {{-{y\over2}P(X_1 = {y\over2}) + yP(X_1 = y)} \over {P(X_1 = {y\over2}) + P(X_1 = y)}}&&\\
&& &=y{{2^n - 2^{n-2}3} \over {2^{n-1}3 + 2^n}}&&\\
&& &= {y\over10} > 0.&&
\end{flalign*}

And, consequently, it is beneficial to switch irrespective of the amount observed. 

\subsection{Continuous variable}

$$f_{X_1}(x_1) = {{1} \over {(x_1 + 1)^2}} \quad \forall \; x_1 > 0.$$

Applying (\ref{qtn:continuous_doubling})

\begin{flalign*}
&&E(B \big\vert Y=y) &= {{-{y\over2}f_{X_1}({y\over2}) + 2y f_{X_1}(y)} \over {f_{X_1}({y\over2}) + 2f_{X_1}(y)}}&&\\
&& &= y{{(y + 2)^2 - (y + 1)^2} \over {(y+2)^2 + 2(y + 1)^2}}&&\\
&& &={{2y^2 + 3y} \over {3y^2 + 8y +6}} > 0.&&
\end{flalign*}

Therefore, always switch irrespective of the amount observed.

%
%
%
%

\section{Extreme values}

The exchange condition that is proposed by \citet[p28]{BK1995} is developed on the \enquote{halving-only} process, since a prior distribution is defined for the larger amount in this instance defined by $X$ and consequently $X_1^\prime = X/2$.

\medskip

While \citet{BK1995} do not distinguish between the initial amounts $X_1$ and $X_1^\prime$ and the final amounts $Y$ and $Z$ it must be realised that $Y=y$ is observed as a result of two distinctly and independent events $c=(x_1;0;0)$ or $c=(2x_1;0,1)$.

\subsection{Discrete variable}

$$P(X_1 = x_1) = p_{k+1} \quad \forall \; x_1 = {2^k}m, k \in \mathbb{Z}.$$

\medskip

Applying (\ref{qtn:discrete_halving}) the expected benefit upon observing $Y=y$ is

\begin{flalign*}
&&E(B \big\vert Y=y) &= {{-{y\over2}P(X_1 = y) + yP(X_1 = 2y)} \over {P(X_1 = y) + P(X_1 = 2y)}}&&\\
&& &=m\cdot{2^k}{{2p_{k+2} - p_{k+1}} \over {2p_{k+2} + 2p_{k+1}}}.&&
\end{flalign*}

The exchange condition is

$$e(y) = 2p_{k+2} - p_{k+1} \quad \forall \, y=m \cdot 2^k, k \in \mathbb{Z}.$$
  
And consequently the switching strategy will be

$$s(y) =\begin{cases}
\text{Switch},& \text{if $e(y) > 0$}\\
\text{Indifference},& \text{if $e(y) = 0$}\\
\text{Do no switch},& \text{if $e(y) < 0$}. 
\end{cases}$$

\bigskip

\subsection{Continuous variable}

\medskip

$$f_{X_1}(x_1) =\begin{cases}
0,&\text{if $x_1 < 1$;}\\
{10^{-2k-1}},&\text{if $10^k \le x_1 < 10^{k+1} \quad \forall \; k \in \mathbb{N}_0$}.
\end{cases}$$

\bigskip

If the content of the allocated envelope is revealed to contain an amount $1/2 \le y < 1$ then the agent can take it that such an instance could only have arisen as a result of the host priming the first envelope with an amount $1 \le x_1 < 2$ and as a result of the \enquote{halving-only} process the second envelope would be assigned an amount $1/2 \le x_1^\prime < 1$. And it is this second envelope that yields the observed amount. No other events are possible for this outcome. The probability of any event $c=(x_1,0,1)$ for all $1 \le x_1 < 2$, with an associated $P(c)=10^{-1}$, implies that the agent should always switch.

\medskip

The agent will now need to consider observing an amount from the interval $10^k \le y < 10^{k+1}$ for all $k \in \mathbb{N}_0$. The agent will need to keep in mind that the amount observed $Y=y$ is possible only as a result of the activities of the host. It is these activities and the resultant outcomes (events) dictate what is observed by the agent. 

\medskip

The agent will noting that, since the \enquote{halving-only} process is operating, consider three possibilities. 

\medskip

\begin{interval}{$10^k \le y < 2 \cdot 10^{k}$}

\medskip

The agent considers observing an amount from the intervals $10^k \le y < 2 \cdot 10^{k}$ for all $k \in \mathbb{N}_0$. 

\medskip

This can be realised for any event $\left\lbrace c \right\rbrace = \left\lbrace x_1,0,0 \right\rbrace $ for all $10^k \le X_1 < 2 \cdot 10^{k}$ or for any event $\left\lbrace c  \right\rbrace = \left\lbrace x_1,0,1 \right\rbrace $ for all $2 \cdot 10^{k} \le X_1 < 4 \cdot 10^{k}$. The probability of any of these events is $P(\left\lbrace c \right\rbrace) = 10^{-2k-1}$ for all $k \in \mathbb{N}_0$. Applying (\ref{qtn:continuous_halving})

 \begin{flalign*}
&&E(B \big\vert Y=y) &= {{-yP(10^k \le X_1 < 2 \cdot 10^{k}) + 4yP(2 \cdot 10^{k} \le X_1 < 4 \cdot 10^{k})} \over
{2P(10^k \le X_1 < 2 \cdot 10^{k}) + 4P(2 \cdot 10^{k} \le X_1 < 4 \cdot 10^{k})}}.&&\\
\intertext{Since, $P(10^k \le X_1 < 2 \cdot 10^{k}) = P(2 \cdot 10^{k} \le X_1 < 4 \cdot 10^{k})$ for all $k \in \mathbb{N}_0$, then} 
&& &= {{-y + 4y} \over {2 + 4}}&&\\
&& &= {y\over2} > 0 \quad \forall k \in \mathbb{N}_0, 10^k \le Y < 2 \cdot 10^{k}.&&
\end{flalign*}

\end{interval}

\medskip

\begin{interval}{$10^{k+1}/2 \le y < 10^{k+1}$}

\medskip

The agent considers observing an amount from the interval $10^{k+1}/2 \le y < 10^{k+1}$.

\medskip

This can be realised for any event $\left\lbrace c \right\rbrace = \left\lbrace x_1,0,0 \right\rbrace \quad \forall \; 10^{k+1}/2 \le X_1 < 10^{k+1}$ with an associated probability $P( \left\lbrace c \right\rbrace ) =10^{-2k-1}$ for all $k \in \mathbb{N}_0$ or for any event $ \left\lbrace c \right\rbrace = \left\lbrace x_1,0,1 \right\rbrace \quad \forall \; 10^{k+1} \le X_1 < 10^{k+1}2$ with an associated probability $P( \left\lbrace c \right\rbrace) = 10^{-2(k+1)-1}$ for all $k \in \mathbb{N}_0$. Applying (\ref{qtn:continuous_halving})

 \begin{flalign*}
&&E(B \big\vert Y=y) &= {{-yP({1\over2}10^{k+1} \le X_1 < 10^{k+1}) + 4yP(10^{k+1} \le X_1 < 2 \cdot 10^{k+1})} \over
{2P({1\over2}10^{k+1} \le X_1 < 10^{k+1}) + 4P(10^{k+1} \le X_1 < 2 \cdot 10^{k+1})}}&&\\
&& &= {{-y{10^{-2k-1}} + 4y{10^{-2(k+1)-1}}} \over {2{10^{-2k-1}} + 4{10^{-2(k+1)-1}}}}&&\\
&& &=-y{{24}\over{51}} < 0 \quad \forall k \in \mathbb{N}_0, {1\over2}10^{k+1} \le Y < 10^{k+1}.&&
\end{flalign*}

\medskip

\end{interval}

\begin{interval}{$10^{k}2 \le y < 10^{k+1}/2$}

\medskip

And finally, the agent considers observing an amount from the interval $10^{k}2 \le y < 10^{k+1}/2$.

\medskip

For any event $\left\lbrace c \right\rbrace = \left\lbrace x_1,0,0 \right\rbrace \quad \forall \; 10^{k}2 \le X_1 < 10^{k+1}/2$ with an associated probability $P(\left\lbrace c \right\rbrace ) =10^{-2k-1}$ for all $k \in \mathbb{N}_0$ or for any event $\left\lbrace c \right\rbrace =(x_1,0,1) \quad \forall \; 10^{k}4 \le X_1 < 10^{k+1}$ with an associated probability $P( \left\lbrace c \right\rbrace ) = 10^{-2k -1}$ for all $k \in \mathbb{N}_0$. Applying (\ref{qtn:continuous_halving})

 \begin{flalign*}
&&E(B \big\vert Y=y) &= {{-yP(2 \cdot 10^{k}2 \le X_1 < {1\over2}10^{k+1}) + 4yP(4 \cdot 10^{k} \le X_1 < 10^{k+1})} \over
{2P(2 \cdot 10^{k} \le X_1 < {1\over2} 10^{k+1}) + 4P(4 \cdot 10^{k} \le X_1 < 10^{k+1})}}&&\\
\intertext{Since, $P(2 \cdot 10^{k} \le X_1 < {1\over2}10^{k+1}) = P(4 \cdot 10^{k} \le X_1 < 10^{k+1})$ for all $k \in \mathbb{N}_0$, then} 
&& &= {{-y + 4y} \over {2 + 4}}&&\\
&& &={y \over 2} > 0 \quad \forall k \in \mathbb{N}_0, 10^{k}2 \le Y < {1\over2}10^{k+1}.&&
\end{flalign*}

\end{interval}

\medskip

We will now consider the instance where $k \in \mathbb{N}_0$. When the agent observes an amount $10^k \le Y < 10^{k+1}$ then in the presence of the \enquote{halving-only} process, what is observed in this interval is only possible if the amount assigned to the first envelope originated from the interval $10^{k} \le X_1 < 10^{k+1}2$. What must now be allowed for is the random allocation of the envelopes. One, of which will reveal the content to be in the interval $10^k \le Y < 10^{k+1}$.

\medskip

If an amount $10^{k+1} \le X_1 < 2 \cdot 10^{k+1}$ is assigned to the first envelope, and as a result of halving the amount $10^{k+1}/2 \le X_1^\prime < 10^{k+1}$ is assigned to the second envelope. It is the content of this second envelope that the agent will be sighting.

\medskip
 
If an amount $10^{k+1}/2 \le X_1 < 10^{k+1}$ is assigned to the first envelope, and as a result of halving the amount $10^{k+1}/4 \le X_1^\prime < 10^{k+1}/2$ is assigned to the second envelope. It is the content of the first envelope that the agent will be sighting.

\medskip

Alternatively, if the first envelope is assigned an amount $10^{k} \le X_1 < 2 \cdot 10^{k}$ and the second $10^{k}/2 \le X_1^\prime < 10^{k}$.

\medskip

And consequently, for all $k \in \mathbb{N}_0$ the agents strategy will be

$$s(y) =\begin{cases}
\text{Switch}, & \text{if ${1\over2} \le y < 1$;}\\
\text{Switch}, & \text{if $10^k \le y < 2 \cdot 10^{k}$;}\\
\text{Switch}, & \text{if $2 \cdot 10^{k} \le Y < {1\over2}10^{k+1}$;}\\
\text{Do not switch}, & \text{if ${1\over2}10^{k+1} \le Y < 10^{k+1}$.}
\end{cases}$$

%
%
%
%

\section{Recurrence relation}

\medskip

\citet{MW2003} questions the findings of \citet{CS2000} and propose the following distribution function associated with a discrete variable. 

\medskip

$$p_n = P(X_1 = 2^n) = {1\over2}(p_{n-1} + 2^{-2n}) = {1\over2}p_{n-1} + 2^{-(2n+1)} \quad \forall \; n \in \mathbb{N}$$
and
$$P(X_1 = 2^0) = p_0 = {1\over12}.$$

Consider

$$p_k = P(X_1 = 2^k) = {1\over2}p_{k-1} + 2^{-(2k+1)}$$

and 

$$p_{k+1} = P(X_1 = 2^{k+1}) = {1\over2}p_k + 2^{-(2k+3)}.$$
 
The \enquote{doubling-only} process will be adopted. Applying (\ref{qtn:discrete_doubling})

\begin{flalign*}
&&E(B \big\vert Y=2^{k+1}) &= {{-{2^k}P(X_1 = 2^k) + 2^{k+1}P(X_1 = 2^{k+1})} \over {P(X_1 = 2^k) + P(X_1 = 2^{k+1})}}&&\\
&& &={{-{2^k}p_k + 2^{k+1}p_{k+1}} \over {p_k + p_{k+1}}}&&\\
&& &={{-{2^k}p_k + 2^{k+1}{({1\over2}p_k + 2^{-(2k+3)})}} \over {p_k + {{1\over2}p_k + 2^{-(2k+3)}}}}&&\\
&& &={{2^{-(k+2)}} \over {{3\over2}p_k + 2^{-(2k+3)}}} &&\\
&& &={{2^{-(k+1)}} \over{3p_k + 2^{-2(k+1)}}} > 0 \quad \forall \, k \in \mathbb{N}_0.&&\\
\end{flalign*}

And consequently it is beneficial to switch irrespective of the amount observed.

\section{Improper distribution}

The mere idea of considering a distribution function that is not normalised would be anathema to many of the authors. As had been documented, only normalised distributions are considered by many authors and used to explain the apparent paradox. At the outset, to consider such distributions may seem unacceptable \citep[p11]{VR2012}. However, the expected benefit conditional on the {sub-$\sigma$-algebra} generated by the observed amount $Y=y$ for a improper distributions are now considered. 

\subsection{Exponential}

For illustrative purposes consider

$$f_{X_1}(x_1) = 2^{-4x^2} \quad \forall \, x>0.$$

Although this function possess similarities to the Weibull it is not normalised and therefore a constant $k$ will be introduced to ensure that it satisfies the criteria of a probability density function (cumulative distribution function).

The probability density function is therefore

$$f_{X_1}(x_1) = k2^{-4x^2} \quad \forall \, x>0.$$

\medskip

A \enquote{doubling-only} process will be adopted, and since $X_1$ is a continuous variable, then applying (\ref{qtn:continuous_doubling})

\begin{flalign*}
&&E(B \big\vert Y=y) &= {{-{y\over2}f_{X_1}({y\over2}) + 2y f_{X_1}(y)} \over {f_{X_1}({y\over2}) + 2f_{X_1}(y)}}&&\\
&& &={{-{y\over2}k2^{-y^2} +2yk2^{-4y^2}} \over {k2^{-y^2} +2k2^{-4y^2}}}.&&
\end{flalign*}

It will be noted that the normalising constant $k$ is factored out and that the exchange condition is

$$e(y) = -{y\over2}2^{-y^2} +2y2^{-4y^2}.$$

From which the strategy associated with the function is as follows:

$$s(y) =\begin{cases}
\text{Switch}, & \text{if $y < \sqrt{2\over3}$;}\\
\text{Indifferent}, & \text{if $y = \sqrt{2\over3}$;}\\
\text{Do not switch}, & \text{if $y > \sqrt{2\over3}$.}
\end{cases}$$

\bigskip

And as expected, this function presents a similar switching strategy to that derived for the Weibull distribution.

\subsection{{Jeffreys' prior}}

The scale invariant \enquote{Jeffreys' prior} $$f_{X_1}(x_1) \propto {1\over x_1}$$ is proposed by \citet[p7]{VR2012} and considered by \citet[p9]{GA2011}. 

\medskip

For illustrative purposes consider

$$f_{X_1}(x_1) = {k \over {x^n}} \quad \forall \, x>0, n \in \mathbb{N}.$$

Although there may be protests from the purists, as before a constant $k$ will be introduced to ensure that it satisfies the criteria of a probability density function (cumulative distribution function).

\medskip

A \enquote{doubling-only} process will be adopted, and since $X_1$ is a continuous variable, then applying (\ref{qtn:continuous_doubling})

\begin{flalign*}
&&E(B \big\vert Y=y) &= {{-{y\over2}f_{X_1}({y\over2}) + 2y f_{X_1}(y)} \over {f_{X_1}({y\over2}) + 2f_{X_1}(y)}}&&\\
&& &=y{{2-{2^{n-1}}}  \over {2 + {2^n}}}.&&
\end{flalign*}

It will be noted that the normalising constant $k$ is factored out, as is $y^{-n}$, and that the exchange condition is

$$e(y) = 2 - 2^{n-1}.$$

From which the strategy associated with the function, irrespective of the observed amount, is as follows:

$$s(y) =\begin{cases}
\text{Switch}, & \text{if $n = 1$;}\\
\text{Indifferent}, & \text{if $n = 2$;}\\
\text{Do not switch}, & \text{if $n > 2$.}
\end{cases}$$

\bigskip

The very idea of an improper distribution may be preposterous to many purists \citep[P39]{SS1997}. While such distribution functions as noted by \citet[p11]{VR2012} may seem absurd, they are, nonetheless, as entertaining as the {two-envelope} problem and should therefore be considered as legitimate.   

%
%
%
%

\section{Summary}

These examples are all legitimate distribution functions and their use to generate an initial amount, although in question by some, do give rise to some interesting and diverse switching strategies. All these strategies are conditional upon the sighting of the amount in the allocated envelope. And essential to any one of the strategies is the content and allocation process adopted by the host.



\def\baselinestretch{1}

\doublespacing

%
%
%

\chapter{The question of a \enquote{paradox}}

\label{chptr:paradox}

\begin{quotation}\small

$\ldots$ thought is the weighing of relative likelihoods of possible events and the act of sampling from the \enquote{posterior}, the probability distribution on unknown events, given the sum total of our knowledge of past events and the present context. If this is so, then the paradigmatic mental object is not a proposition, standing in all its eternal glory with its truth value emblazoned on its chest, but the random variable $x$, its value subject to probabilities but still not fixed $\ldots$. The simplest example where human thinking is clearly of this kind may well be the case where probabilities can be made explicit: gambling. Here we are quite conscious that we are weighing likelihoods (and even calculating them if we are mathematically inclined). If we accept this, the division of mathematics corresponding to this realm of experience is not logic but probability and statistics.

\begin{flushright}
--- The Dawning of the Age of Stochasticity \citep[5]{MD1999}
\end{flushright}

\end{quotation}

\medskip

The two-envelope paradox has generated much debate within academic circles from diverse disciplines such as mathematics, statistics, economics, philosophy, and psychology. The perception of this research is that irrespective of which envelope is allocated to the agent, it would always be beneficial for the agent to switch to the complementary envelope.

\medskip

Although \citet[191]{SR1992} dismisses the use of probability to explain the paradox, he still uses the words: \enquote{probability}, \enquote{equal probability}, \enquote{chances}, \enquote{$\ldots$ the chances of gaining or losing are equal}, and \enquote{$\ldots$ odds}. All these terms and expressions have probabilistic connotations and, subsequently, what is considered to be devoid of probability must be resolved by a embracing a probabilistic argument. \citet{SP2010}, by contrast, argues that since the two-envelope problem is a decision theoretic, the agent may assume --- in the absence of sighting the content of the allocated envelope --- that since the contents are fixed at the time of selection, that they remain unchanged during the process and, therefore, that it is irrelevant whether he or she chooses to switch or not. Whether sighting the content provides any information to assist the agent in deciding to switch or not is dependent on the information provided by the host of the game.

\medskip

It is true that the content of one envelope is twice that of the other. It is also true that relative to the content of the first envelope --- if the halving or doubling process was adopted by the host --- the second envelope is either half or double that of the first. However, based on what has been discussed in the previous chapters, this can be considered to be only half the story. The allocation of either envelope to the agent is reliant on the outcome of a random event, and in this instance, there are two possibilities: either the event $(x_1; \omega_2; 0)$ or the event $(x_1; \omega_2; 1)$ has been realised by the host. These events may not have occurred with equal likelihood as revealed in Chapter \ref{chptr:strategies}.

\section{A likely outcome}

\begin{quote}\small

\enquote{Probability is now irrelevant,} said the Sorcerer.

\begin{flushright}
--- Satan, Cantor, and Infinity \citep[191]{SR1992}
\end{flushright}

\end{quote}

Not so! We say that the outcome of a toss of an unbiased coin is either a head or a tail. It is also said that the content of the other envelope is either $2n$ or $n/2$. As stated by \citet{MD1999} such events are best described in the terms of probability.

\medskip

Before explaining the reasoning behind the apparent paradox, it is necessary to introduce some notation. Where possible, all arguments are expressed in terms of the realisations of the experiments performed by the host; consequently, random variables are only used when necessary in order to align with the thinking of Smullyan and to argue in favour of probability. 

\medskip

The amount assigned to the second envelope is dependent upon the outcome of an experiment performed by the host. The factor $g(\omega_2)$ is derived from the outcome of tossing an unbiased coin:  

$$g(\omega_2) = {1\over2}(1 - \omega_2) + 2\omega_2 = 
\begin{cases}
{1\over2},&\text{if $\omega_2 = 0$;}\\
2,&\text{if $\omega_2 = 1$.}
\end{cases}$$

The random selection of one of the primed (indistinguishable) envelopes and the allocation of that envelope to the agent is represented by the symmetric matrix $\textbf{A}(\omega_3)$. 

$$\textbf{A}(\omega_3) =
\begin{bmatrix}
1-\omega_3 & \omega_3\\
\omega_3 & 1 - \omega_3
\end{bmatrix} = \begin{cases}
I,&\text{if $\omega_3 = 0$;}\\
J,&\text{if $\omega_3 = 1$.}
\end{cases}$$

where $I$ and $J$ are the identity and exchange matrices, respectively.

For the outcomes of the experiments performed by the host, namely $c = (x_1; \omega_2; 0)$ --- allocate the first envelope --- and $c = (x_1; \omega_2; 1)$ --- allocate the second envelope, the allocated and complementary envelope contents are as follows: 

\begin{flalign}
&&\begin{matrix}\textbf{C}({c}) \end{matrix} &= \begin{matrix}\textbf{C}(x_1;\omega_2;\omega_3) \end{matrix} &&\\ 
&& &= 
\begin{matrix} x_1 \end{matrix} 
\begin{bmatrix}
1 & g(\omega_2)
\end{bmatrix}
\begin{matrix} \textbf{A}(\omega_3)
\end{matrix}&&\notag\\
&& &=\begin{cases}
\begin{matrix} x_1 \end{matrix}
\begin{bmatrix}
1 & g(\omega_2)
\end{bmatrix},&\text{if $\omega_3 = 0$;}\\
\begin{matrix} x_1 \end{matrix} \begin{bmatrix}
g(\omega_2) & 1
\end{bmatrix},&\text{if $\omega_3 = 1$.}
\end{cases}\label{qtn:allocation}
\end{flalign}

Therefore, the benefit of switching, given the event $(x_1; \omega_2; \omega_3)$, is 
    
\begin{flalign}
b(x_1;\omega_2;\omega_3) = \begin{cases}
x_1(g(\omega_2) - 1),&\text{if $\omega_3 = 0$;}\\
x_1(1 - g(\omega_2),&\text{if $\omega_3 = 1$.}
\end{cases}\label{qtn:benefit_1}
\end{flalign}

%
%
%

\section{The Story}

\begin{quote}\small

There are two sealed envelopes on the table. You are told that one of them contains twice as much money as the other. You pick one of them and open it to see how much money is inside. Let's say you find $\$ 100$ in it. Then you are given the option of keeping it or trading it for the other envelope. Now, the other envelope has twice as much or half as much with equal probability. Thus, the chances are equal that the other envelope has $\$ 200$ or $\$ 50$, and the chances of your gaining or losing are equal. And so the odds are in your favour if you trade.

\begin{flushright}
--- Satan, Cantor, and Infinity \citep[189]{SR1992}
\end{flushright}

\end{quote}

As mentioned in the literary review the importance of posing the two-envelope problem in a form that supports a probabilistic argument is pivotal for the derivation of a switching strategy. The propositions presented by \citet[190]{SR1992} are repeated here for convenience:

\medskip

\begin{proposition}\label{prpstn:one}
The amount that you will gain, if you do gain, is greater than the amount you will lose, if you do lose.
\end{proposition}

\begin{proposition}\label{prpstn:two}
The amounts are the same.
\end{proposition}

\medskip

For the event $(x_1; \omega_2; 0)$, the agent would be allocated the envelope containing the amount $x_1$, while the complementary envelope would contain the amount $x_1g(\omega_2)$. Consequently, if the agent were to switch envelopes, the benefit  would be $x_1(g(\omega_2) - 1)$. The agent should argue similarly for the event $(x_1; \omega_2; 1)$.
 
\medskip

It is clear from Equation \ref{qtn:benefit_1}, in the absence of any information, that  $b(x_1; \omega_2; 0) = -b(x_1; \omega_2; 1)$ supports Proposition~\ref{prpstn:two}. In the words of \citet[191]{SR1992}: \enquote{If you gain on the trade, you gain $d$ dollars, and if you lose on the trade, you lose $d$ dollars. And so the amounts are the same after all}.

\medskip

This is akin to the agent having no information about the game and consequently includes the instance where the agent is not afforded the opportunity of sighting the content of the allocated envelope before deciding to switch envelopes or not.

\section{Half the story}

\citet[p190]{SR1992} has in Proposition~\ref{prpstn:two} chosen to present the equally likely events $(x_1; \omega_2; 0)$ and $(x_1; \omega_2; 1)$ and then demonstrates the outcome by way of an example that the benefits of trading amounts are the same. 

\medskip

Suppose, however that the host identifies an amount $x_1$ and after assigning this amount to the first envelope immediately allocates the envelope to the agent. The host then identifies a subsequent amount $x_1g(\omega_2)$ and then assigns this to the second (complementary) envelope. 

\medskip

The only possible event representing the outcome in this instance is $(x_1; \omega_2; 0)$, which reflects the allocation to the agent of the envelope that contains $x_1$. The complementary envelope will then contain $x_1g(\omega_2)$, which can take the value $x_1/2$ or $2x_1$, depending on $\omega_2$. The \enquote{halving or doubling} process is in operation here, since a \enquote{doubling-only} process would result in $2x_1$ only and consequently there is no gamble. In other words, the agent knows for certain that the complementary envelope contains twice the amount of the allocated envelope, and therefore a \enquote{doubling-only} process is not possible.

\medskip

For the event $c = (x_1; \omega_2; 0)$, $x_1$ is allocated to the agent, while the amount in the complementary envelope could be either ${x_1}/2$ or $2x_1$ with an equal likelihood. Hence, if the agent were to switch, he or she would gain $x_1$ or lose ${x_1}/2$. This is the reasoning behind Proposition~\ref{prpstn:one}. It can be argued that a rational agent would switch envelopes if the certain amount $x_1$ was considered against the pay-off associated with a bet on either ${x_1}/2$ or $2x_1$ \citep[ p. 212, 216]{CJ1992}.  

\medskip
  
A risk-neutral and rational agent should argue that for any gamble that the average payoff of the complementary envelope being greater than the allocated envelope supports the decision to trade. However, in this way no consideration has been paid to the content and allocation process, or more specifically the random selection and allocation of the envelopes. However, to trade and then resort to the same argument is fallacious, since the resultant envelopes amounts to the realisation of the event $(x_1; \omega_2; 1)$, and the observed content of the second envelope is now not the amount $x_1g(\omega_2)$, which is a nebulous amount, but either $x_1/2$ or $2x_1$. Neither of these amounts can be observed simultaneously. Thus, had the agent observed $x_1/2$, he or she would infer that the complementary envelope contained either $x_1/4$ or $x_1$. In the absence of information, whether having sight of the contents or not, the agent should be indifferent to switching envelopes or not. The benefit of switching, given the event $(x_1; \omega_2; 0)$, is therefore    

\begin{flalign}
b(x_1;\omega_2;0) =  x_1 (g(\omega_2) - 1) = \begin{cases}
{-{1\over 2}x_1},&\text{if $\omega_2 = 0$;}\\
x_1,&\text{if $\omega_2 = 1$.}
\end{cases}\label{qtn:benefit_2}
\end{flalign}

and since these outcomes are equally likely, the agent should switch envelopes.

%
%
%

\section{The full story}

\citet{SR1992} allows for the events $(x_1; \omega_2; 0)$ and $(x_1; \omega_2; 1)$ when discussing Proposition 2, but only the event $(x_1; \omega_2; 0)$ when discussing Proposition 1. Ignoring the event $(x_1; \omega_2; 1)$ when discussing Proposition 1 amounts to dismissing the random allocation of the envelopes and, consequently, the problem becomes a gamble for which the payoff is the difference between the expected content of the complementary envelope and the sighted content of the allocated envelope.

\medskip

Possibility \ref{pssblty:three} and Possibility \ref{pssblty:four} Before the agent is afforded the opportunity of sighting the content of the allocated envelopes, he or she should consider that the events $(x_1; \omega_2; 0)$ and $(x_1; \omega_2; 1)$ are equally likely. For the event $(x_1; \omega_2; 0)$, the agent is allocated the envelope that contains $x_1$ and the complementary envelope therefore contains either ${x_1}/2$ or $2x_1$ with an equal probability. The agent can, in this instance, after sighting the content \citep[428]{CS2000}, correctly argue that the complementary envelope contains either half or double the content of allocated envelope. Unfortunately, the same reasoning cannot be applied to the event $(x_1; \omega_2; 1)$. In this instance, the content of the allocated envelope will be either ${x_1}/2$ or $2x_1$ and the complementary envelope will contain $x_1$. Upon sighting the content of the allocated envelope, if it were ${x_1}/2$ the agent would erroneously reason that the content of the complementary envelope would be either ${x_1}/4$ or $x_1$, of which only $x_1$ is possible. Similarly, if the agent were to sight the amount $2x_1$, then using the same reasoning he or she would expect the complementary envelope to contain either $x_1$ or $4x_1$, of which only $x_1$ is correct.

\medskip

The switching decision is not dependent on the expected content of the allocated envelope but rather on the sighted content of the allocated envelope and the expected content of the complementary envelope. It is therefore necessary for the agent --- upon sighting the content --- to apply one of the strategies discussed in Chapter~\ref{chptr:strategies}.

\medskip

Moreover, the consideration by \citet[158]{CJ2002} of equally likely events $(x_1; \omega_2; 0)$  and $(x_1; \omega_2; 1)$ is necessary when comparing Proposition~\ref{prpstn:one} with Proposition~\ref{prpstn:two}.

%
%
%

\section{Summary}

\begin{quote}\small

There is an ample supply of misleading leads, false clues, and dead ends to confuse the issue $\ldots$ you will be dealing with concepts that are central to a number of philosophical doctrines or controversies. Not only are the fundamental notions of decision theory involved, --- preferences, desirability, probability, possibility, and expected utility --- but one also finds the slightly suspect Kripkean idea of fixing the reference of a name by means of a description.

\begin{flushright}
--- The Mystery of Julius: A Paradox in Decision Theory \citep[5]{CC1995}
\end{flushright}

\end{quote}

\medskip
 
The introduction of the concept of a probability triple $(\Omega,\mathcal{F},P)$ and the defined sample space follows the approach put forward by \citet{CR1992}, who explore the paradox from a frequentist perspective. In their proposal, they adopt only the doubling strategy. \citet{BF1996} acknowledges that the paradox is a fallacy associated with the failure to define the appropriate sample space. He concurs with \citet{CR1992} in their frequentist and Bayesian explanation.


It is not correct to argue that the complementary envelope $Z$ contains $y/2$ or $2y$ in the absence of the stated information. The events that originate from the activities of the host are essential for the agent to know. In the absence of this information, sighting the content  or not has no bearing on the expected outcome and, consequently, the agent should be indifferent about switching or not. 

\medskip

A fundamental principle for the resolution of any problem is its re-formulation into a format that lends itself readily to the application of methods that may realise an outcome that was not initially obvious. In the absence of such an initiative, academic sleight-of-hand has gone unnoticed. There has thus been the realisation that there is more to this simple problem than meets the eye.  

\medskip

The random variable $Y$, which represents the content of the allocated envelope according to current and previous thinking, will always have the value $Y=y$ and should be treated as fixed. 

\citet{SR1992} attempts to explain the paradox using logic; however in Proposition~\ref{prpstn:one} he fails to consider that he was not considering the benefit as he described in Proposition~\ref{prpstn:two} and, therefore, that any comparison would be impossible. However, had he considered expressing Proposition~\ref{prpstn:one} as expressed herein, he would have realised that Proposition~\ref{prpstn:one} and Proposition~\ref{prpstn:two} actually agree.

\bigskip

There are two possibilities. The agent is allocated the first envelope containing the initial amount $X_1 = x_1$ and the second envelope is then assigned the amount $X_1^\prime = {x_1}/2$ or  the agent is allocated the first envelope containing the initial amount $X_1 = x_1$ and the second envelope is then assigned the amount $X_1^\prime = 2x_1$.

\medskip

Whether sighting the content or not, in the absence of all other information, Proposition~\ref{prpstn:two}  is true. In sighting the content of the allocated envelope, Proposition~\ref{prpstn:one} is not true since only the certain outcome of a specific event is considered.




\def\baselinestretch{1}

\chapter{Conclusion}

\def\baselinestretch{1.66}

\begin{quote}\small

Today's scientists have substituted mathematics for experiments, and they wander off through equation after equation, and eventually build a structure which has no relation to reality.

\begin{flushright}
--- Nikola Tesla
\end{flushright}

\end{quote}

Does the two-envelope paradox exist? This question, which has entertained, challenged, and frustrated many researchers, has received much attention since the original formulation by \citet[p133]{KM1943}. In all its variants, however, it remains essentially the same, and therefore the overriding question remains: Is there a benefit to switching? While a number of approaches have proposed to resolve this dilemma, the approach adopted in this paper has revealed a missing piece of the puzzle.

\medskip

The apparent two-envelope paradox exists only when the problem is formulated from an incomplete or inadequate description of the realisable benefits and when there is complete disregard for some of the fundamental and essential issues associated with probability. For example, \citet{NJ1998} formulates elaborate solutions to a simple problem, and then ignores the fundamental concepts essential for any mathematically derived solution. In conclusion, provided there is no knowledge of the distribution function from which the monetary amounts were drawn or the limits of the monetary content in each of the envelopes, and irrespective of whether the content of the chosen envelope is sighted or not, there is no benefit to the agent of switching.

\medskip

This thesis has demonstrated that the benefit of switching therefore does not rely upon a single event but rather upon the statistical expectation associated with the activities of the host for the sighted content of the allocated envelope. Because the posing of the problem is set in probabilistic formalism, it lends itself readily to statistical and probabilistic resolution methods. The activities of the host, when revealed to the agent, are sufficient for the agent to derive an envelope-switching strategy. In the absence of this information, the agent should be indifferent to switching.

\medskip

Much of what is purported to resolve the problem (and thereby the paradox) results from poor formulation and incorrect structuring as well as from the outright ignorance of fundamental statistical methods. Oftentimes, the problem is contextualised in unnecessary jargon and complex mathematical methods. The paradox that has long possessed academics has no legitimacy, and only exists for those that choose it to. Indeed, the substantiation of the paradox stems from the flawed interpretation of an ill-posed problem and an abuse of basic statistical methods.

\medskip

The examples, where sighting the content of the allocated envelope, have revealed that in some instances it is always beneficial to switch. There are instances where it is not beneficial to switch and there are others where the strategy is mixed. In all instances the expected benefit is conditional on the {sub-$\sigma$-algebra} induced by the observed amount represented by the random variable $Y$, irrespective if the observed amount is a probability zero event. Arguments questioning the relevance of distributions or functions over infinite amounts are no longer valid. Distribution or functions that represent infinite amounts are legitimate and in the presence of the necessary information it will be realised that the approach derived in this paper, that exploits the features of the {sub-$\sigma$-algebra}, can yield an appropriate switching strategy. As noted by \citet[p4]{MW2003} switching envelopes is conditional upon the sighting of a particular value, in the allocated envelope, and not whether swapping is better \enquote{on average} over repeated trials. In instances where the \enquote{exchange condition} is positive for all observed amounts, implying that it is not necessary to view the contents and that it would make sense to switch envelopes irrespective of sighting the content or not does not legitimise the paradox \citep[p28]{BK1995}.

\medskip

\section{In closing}

\begin{quote}\small

$\ldots$ statistical information on frequencies within a large, repetitive class of events is strictly irrelevant to a decision whose outcome depends on a single trial.

\begin{flushright}
--- Risk, Ambiguity, and the Savage Axioms \citep[2]{ED1961}
\end{flushright}

\end{quote}

In the \enquote{halving or doubling} process the traditional argument is to assess the benefit of switching the allocated envelope for the complementary envelope. The content of one of the envelopes, for this process, is acknowledged to be either half or double the other. With this information the agent is assumed to reason that there are two amounts for consideration and that the conditional probability of these amounts are all that are necessary to derived the expected benefit. But, as has been demonstrated the conditional expectation for this content and allocation process is dependent on the host sampling three distinct amounts. These possible initial amounts are all induced by the amount observed by the agent. 

\medskip

For any play of the game the agent may lose or the agent may win. Alternatively, if the agent is mathematically inclined, he or she may apply likelihoods to each of the induced outcomes and derive an expected benefit of switching envelopes.

\section{The paradox $\ldots$} 

\vfill\eject

\begingroup

\topskip0pt
\vspace*{\fill}

\begin{quote}\small

This is your last chance. After this, there is no turning back. You take the blue pill -- the story ends, you wake up in your bed and believe whatever you want to believe. You take the red pill -- you stay in Wonderland, and I show you how deep the rabbit hole goes. 

\begin{flushright}
--- Morpheus (The Matrix)
\end{flushright}

\end{quote}

\vspace*{\fill}

\endgroup

\backmatter 







\bibliographystyle{plainnat}

\renewcommand{\bibname}{Reference} 

\bibliography{one} 

\begin{thebibliography}{41}
\providecommand{\natexlab}[1]{#1}
\providecommand{\url}[1]{\texttt{#1}}
\expandafter\ifx\csname urlstyle\endcsname\relax
  \providecommand{\doi}[1]{doi: #1}\else
  \providecommand{\doi}{doi: \begingroup \urlstyle{rm}\Url}\fi

\bibitem[Albers et~al.(2005)Albers, Kooi, and Schaafsma]{AK2005}
C.~J. Albers, B.~P. Kooi, and W.~Schaafsma.
\newblock Trying to resolve the {Two-Envelope} problem.
\newblock \emph{Synthese}, 145\penalty0 (1):\penalty0 89--109, May 2005.
\newblock \doi{10.1007/s11229-004-7665-5}.
\newblock URL \url{http://dx.doi.org/10.1007/s11229-004-7665-5}.

\bibitem[Ash(1972)]{AR1972}
R.~B. Ash.
\newblock \emph{Real Analysis and Probability}.
\newblock Academic Pr, June 1972.
\newblock ISBN 0120652404.

\bibitem[Athreya and Lahiri(2010)]{AL2010}
K.~B. Athreya and S.~N. Lahiri.
\newblock \emph{Measure Theory and Probability Theory}.
\newblock Springer, softcover reprint of hardcover 1st ed. 2006 edition,
  November 2010.
\newblock ISBN 1441921915.

\bibitem[Blachman and Kilgour(2001)]{BK2001}
Nelson~M. Blachman and D.~Marc Kilgour.
\newblock Elusive optimality in the box problem.
\newblock \emph{Mathematics Magazine}, 74\penalty0 (3):\penalty0 pp. 171--181,
  2001.
\newblock ISSN 0025570X.
\newblock URL \url{http://0-www.jstor.org.oasis.unisa.ac.za/stable/2690718}.

\bibitem[Brams and Kilgour(1995)]{BK1995}
S.~J. Brams and D.~M. Kilgour.
\newblock The box problem: To switch or not to switch.
\newblock \emph{Mathematics Magazine}, 68\penalty0 (1):\penalty0 27--34, 1995.
\newblock ISSN 0025570X.
\newblock URL \url{http://www.jstor.org/stable/2691373}.

\bibitem[Broome(1995)]{BJ1995}
J.~Broome.
\newblock The two-envelope paradox.
\newblock \emph{Analysis}, 55\penalty0 (1):\penalty0 6--11, 1995.
\newblock ISSN 00032638.
\newblock URL \url{http://www.jstor.org/stable/3328613}.

\bibitem[Bruss(1996)]{BF1996}
F.~T. Bruss.
\newblock The fallacy of the two envelopes problem.
\newblock \emph{Math. Sci.}, 21\penalty0 (2):\penalty0 112--119, 1996.
\newblock ISSN 0312-3685.

\bibitem[Bruss and Ruschendorf(2000)]{BR2000}
F.~T. Bruss and L.~Ruschendorf.
\newblock The switching problem and conditionally specified distribution.
\newblock \emph{Math. Sci.}, 25\penalty0 (1):\penalty0 47--53, 2000.
\newblock ISSN 0312-3685.

\bibitem[Cargile(1992)]{CJ1992}
J.~Cargile.
\newblock On a problem about probability and decision.
\newblock \emph{Analysis}, 52\penalty0 (4):\penalty0 211--216, 1992.
\newblock URL \url{http://www.jstor.org/stable/3328337.html}.

\bibitem[Castell and Batens(1994)]{CB1994}
P.~Castell and D.~Batens.
\newblock The two envelope paradox: The infinite case.
\newblock \emph{Analysis}, 54\penalty0 (1):\penalty0 pp. 46--49, 1994.
\newblock ISSN 00032638.
\newblock URL \url{http://0-www.jstor.org.oasis.unisa.ac.za/stable/3328103}.

\bibitem[Chase(2002)]{CJ2002}
J.~Chase.
\newblock {The non-probabilistic two envelope paradox}.
\newblock \emph{Analysis}, 62\penalty0 (2):\penalty0 157--160, 2002.
\newblock \doi{10.1093/analys/62.2.157}.
\newblock URL \url{http://analysis.oxfordjournals.org}.

\bibitem[Chen(2007)]{CJ2007}
G.~J. Chen.
\newblock The puzzle of the {Two-Envelope} puzzle.
\newblock \emph{{SSRN} {eLibrary}}, February 2007.
\newblock URL \url{http://papers.ssrn.com/sol3/papers.cfm?abstract_id=1132506}.

\bibitem[Chihara(1995)]{CC1995}
C.~S. Chihara.
\newblock The mystery of {Julius}: A paradox in decision theory.
\newblock \emph{Philosophical Studies: An International Journal for Philosophy
  in the Analytic Tradition}, 80\penalty0 (1):\penalty0 1--16, October 1995.
\newblock URL \url{http://www.jstor.org/stable/4320617.html}.

\bibitem[Christensen and Utts(1992)]{CR1992}
R.~Christensen and J.~Utts.
\newblock Bayesian resolution of the \enquote{Exchange Paradox}.
\newblock \emph{The American Statistician}, 46\penalty0 (4):\penalty0 274--276,
  1992.
\newblock ISSN 00031305.
\newblock URL \url{http://www.jstor.org/stable/2685310}.

\bibitem[Clark and Shackel(2000)]{CS2000}
M.~Clark and N.~Shackel.
\newblock {The two-envelope paradox}.
\newblock \emph{Mind}, 109\penalty0 (435):\penalty0 415--442, 2000.
\newblock \doi{10.1093/mind/109.435.415}.
\newblock URL
  \url{http://mind.oxfordjournals.org/cgi/content/abstract/109/435/415}.

\bibitem[de~Saint-Exup\'ery(1945)]{AS1943}
A.~de~Saint-Exup\'ery.
\newblock \emph{The Little Prince}.
\newblock Egmont UK limited, 1945.

\bibitem[Ellsberg(1961)]{ED1961}
D.~Ellsberg.
\newblock Risk, ambiguity, and the savage axioms.
\newblock \emph{The Quarterly Journal of Economics}, 75\penalty0 (4):\penalty0
  643--669, November 1961.
\newblock ISSN 00335533.
\newblock \doi{10.2307/1884324}.
\newblock URL
  \url{http://www.jstor.org/discover/10.2307/1884324?uid=2129&uid=2&uid=70&uid=4&sid=21101691478853}.
\newblock {ArticleType:} primary\_article / Full publication date: Nov., 1961 /
  Copyright 1961 The {MIT} Press.

\bibitem[Gill(2011)]{GA2011}
R.~D. Gill.
\newblock \emph{Anna Karenina and The Two Envelopes Problem (Fourth, still very
  incomplete, draft)}.
\newblock 2011.
\newblock URL \url{http://www.math.leidenuniv.nl/~gill/tep.pdf}.

\bibitem[Jackson et~al.(1994)Jackson, Menzies, and Oppy]{JF1994}
F.~Jackson, P.~Menzies, and G.~Oppy.
\newblock The two envelope \enquote{Paradox}.
\newblock \emph{Analysis}, 54\penalty0 (1):\penalty0 43--45, 1994.

\bibitem[Jeffrey(1986)]{JR1986}
R.~C. Jeffrey.
\newblock Judgmental probability and objective chance.
\newblock \emph{Erkenntnis}, 24\penalty0 (1):\penalty0 5--16, 1986.
\newblock ISSN 0165-0106.
\newblock \doi{10.1007/BF00183197}.
\newblock URL \url{http://dx.doi.org/10.1007/BF00183197}.

\bibitem[Katz and Olin(2007)]{KO2007}
B.~D. Katz and D.~Olin.
\newblock A tale of two envelopes.
\newblock \emph{Mind}, 116\penalty0 (464):\penalty0 pp. 903--926, 2007.
\newblock ISSN 00264423.
\newblock URL \url{http://0-www.jstor.org.oasis.unisa.ac.za/stable/30166515}.

\bibitem[Kraitchik(1942)]{KM1943}
M.~Kraitchik.
\newblock \emph{Mathematical Recreations}.
\newblock W. W. Norton \& Company, Inc., New York (2nd revised edition, 2006,
  Dover Publications, New York), 1942.
\newblock ISBN 0-486-45358-8.

\bibitem[Loredo(2004)]{LT2004}
T.~Loredo.
\newblock Critical thinking, summer 2004: The two-envelope paradox.
\newblock 2004.
\newblock URL
  \url{http://www.astro.cornell.edu/staff/loredo/bayes/two-envelope.pdf}.

\bibitem[McDonnell and Abbott(2009)]{MA2009}
M.~D. McDonnell and D.~Abbott.
\newblock Randomized switching in the two-envelope problem.
\newblock \emph{Proceedings of the Royal Society A: Mathematical, Physical and
  Engineering Science}, 2009.
\newblock \doi{10.1098/rspa.2009.0312}.
\newblock URL
  \url{http://rspa.royalsocietypublishing.org/content/early/2009/07/31/rspa.2009.0312.abstract}.

\bibitem[McGrew et~al.(1997)McGrew, Shier, and Silverstein]{MC1997}
T.~J. McGrew, D.~Shier, and H.~S. Silverstein.
\newblock The two-envelope paradox resolved.
\newblock \emph{Analysis}, 57\penalty0 (1):\penalty0 28--33, 1997.

\bibitem[Meacham and Weisberg(2003)]{MW2003}
C.~Meacham and J.~Weisberg.
\newblock Clark and {Shackel} on the {Two-Envelope} paradox.
\newblock http://mind.oxfordjournals.org/cgi/pdf\_extract/112/448/685, October
  2003.
\newblock URL \url{http://mind.oxfordjournals.org/cgi/pdf_extract/112/448/685}.

\bibitem[Mood et~al.(1974)Mood, Graybill, and Boes]{MA1974}
A.~M. Mood, F.~A. Graybill, and D.~C. Boes.
\newblock \emph{Introduction to the theory of statistics}.
\newblock McGraw-Hill, third edition, 1974.
\newblock ISBN 0-07-042864-6.

\bibitem[Mumford(1999)]{MD1999}
D.~Mumford.
\newblock The dawning of the age of stochasticity, 1999.
\newblock URL
  \url{http://www.dam.brown.edu/people/mumford/Papers/OverviewPapers/DawningAgeStoch.pdf}.

\bibitem[Nalebuff(1989)]{NB1989}
B.~Nalebuff.
\newblock The other person's envelope is always greener.
\newblock \emph{Journal of Economic Perspectives}, 3\penalty0 (1):\penalty0
  171--81, Winter 1989.
\newblock URL \url{http://ideas.repec.org/a/aea/jecper/v3y1989i1p171-81.html}.

\bibitem[Norton(1998)]{NJ1998}
J.~Norton.
\newblock Where the sum of our expectation fails us: The exchange paradox.
\newblock \emph{Pacific Philosophical Quarterly}, 79\penalty0 (1):\penalty0
  34--58, 1998.
\newblock ISSN 1468-0114.
\newblock \doi{10.1111/1468-0114.00049}.
\newblock URL \url{http://dx.doi.org/10.1111/1468-0114.00049}.

\bibitem[Ok(2007)]{NY1234}
E.A. Ok.
\newblock \emph{Real analysis with economic applications}.
\newblock Princeton University Press, 2007.
\newblock URL \url{https://files.nyu.edu/eo1/public/Book-PDF/pChapterIII.pdf}.

\bibitem[Priest and Restall(2007)]{PR2007}
G.~Priest and G.~Restall.
\newblock Envelopes and indifference, 2007.
\newblock URL \url{http://philpapers.org/rec/PRIEAI}.

\bibitem[Rawling(1994)]{RP1994}
P.~Rawling.
\newblock A note on the two envelopes problem.
\newblock \emph{Theory and Decision}, 36\penalty0 (1):\penalty0 97--102, 1994.
\newblock ISSN 0040-5833.
\newblock \doi{10.1007/BF01075299}.
\newblock URL \url{http://dx.doi.org/10.1007/BF01075299}.

\bibitem[R\'enyi(1970)]{RA1970}
A.~R\'enyi.
\newblock \emph{Foundations of Probability}.
\newblock San Francisco: Holden-Day, 1970.

\bibitem[Scott and Scott(1997)]{SS1997}
A.~D. Scott and M.~Scott.
\newblock What's in the two envelope paradox?
\newblock \emph{Analysis}, 57\penalty0 (1):\penalty0 pp. 34--41, 1997.
\newblock ISSN 00032638.
\newblock URL \url{http://0-www.jstor.org.oasis.unisa.ac.za/stable/3328432}.

\bibitem[Silver(1994)]{SE1994}
E.~A. Silver.
\newblock On mathematical problem posing.
\newblock \emph{For the Learning of Mathematics}, 14\penalty0 (1):\penalty0 pp.
  19--28, 1994.
\newblock ISSN 02280671.
\newblock URL \url{http://0-www.jstor.org.oasis.unisa.ac.za/stable/40248099}.

\bibitem[Smullyan(1992)]{SR1992}
R.~M. Smullyan.
\newblock \emph{Satan, Cantor, And Infinity and Other {Mind-Boggling} Puzzles}.
\newblock Knopf, first printing edition, November 1992.
\newblock ISBN 0679406883.

\bibitem[Sobel(1994)]{SJ1994}
J.~H. Sobel.
\newblock Two envelopes.
\newblock \emph{Theory and Decision}, 36\penalty0 (1):\penalty0 69--96, 1994.

\bibitem[{Spanos}(2013)]{SA2013}
A.~{Spanos}.
\newblock {The Two Envelope Problem: a Paradox or Fallacious Reasoning?}
\newblock \emph{ArXiv e-prints}, January 2013.

\bibitem[Sutton(2010)]{SP2010}
P.~A. Sutton.
\newblock The epoch of incredulity: A response to {Katz} and {Olin\rq s} \lq {A
  Tale of Two Envelopes}\rq.
\newblock \emph{Mind}, 119\penalty0 (473):\penalty0 159--169, 2010.
\newblock \doi{10.1093/mind/fzp164}.
\newblock URL
  \url{http://mind.oxfordjournals.org/content/119/473/159.abstract}.

\bibitem[{Vazquez}(2012)]{VR2012}
R.~A. {Vazquez}.
\newblock {The two envelopes probability paradox: Much ado about nothing}.
\newblock \emph{ArXiv e-prints}, June 2012.

\end{thebibliography}


\end{document}